\documentclass[
    aps,
    prd,
    reprint,
    nolongbibliography,
    nobibnotes,
    nofootinbib,
    amssymb,
    amsmath,
]{revtex4-2}

\usepackage{hyperref}
\hypersetup{hidelinks}

\usepackage[dvipsnames]{xcolor}
\usepackage{orcidlink}
\usepackage{physics}
\usepackage{xspace}
\usepackage{soul}

\newcommand{\Lhat}{\hat{\mathcal L}}

\newcommand{\LIGO}{\affiliation{LIGO Laboratory, Massachusetts Institute of Technology, Cambridge, MA 02139, USA}}
\newcommand{\kavli}{\affiliation{Kavli Institute for Astrophysics and Space Research, Massachusetts Institute of Technology, Cambridge, MA 02139, USA}}
\newcommand{\MIT}{\affiliation{Department of Physics, Massachusetts Institute of Technology, Cambridge, MA 02139, USA}}

\newcommand{\TOPclr}{green\xspace}
\newcommand{\LOGclr}{purple\xspace}
\newcommand{\locclr}{orange\xspace}

\newcommand{\TOPIMFalpha}{\ensuremath{0.44^{+1.15}_{-1.47}}\xspace}
\newcommand{\LOGIMFalpha}{\ensuremath{0.84^{+0.53}_{-0.46}}\xspace}
\newcommand{\TOPIMFbeta}{\ensuremath{1.08^{+1.72}_{-1.85}}\xspace}
\newcommand{\LOGIMFbeta}{\ensuremath{0.93^{+1.82}_{-1.66}}\xspace}

\newcommand{\TOPSFRDten}{\ensuremath{-9.95^{+1.32}_{-1.39}}\xspace}
\newcommand{\TOPSFRDsixteen}{\ensuremath{-10.61^{+1.40}_{-1.28}}\xspace}
\newcommand{\LOGSFRDten}{\ensuremath{-10.92^{+0.99}_{-1.23}}\xspace}
\newcommand{\LOGSFRDsixteen}{\ensuremath{-8.40^{+0.56}_{-0.61}}\xspace}

\newcommand{\LOGMRDten}{\ensuremath{1.73^{+0.55}_{-0.42}}\xspace}
\newcommand{\TOPMRDten}{\ensuremath{0.93^{+0.61}_{-0.51}}\xspace}
\newcommand{\LOGlocMRDten}{\ensuremath{4.32^{+1.43}_{-1.12}}\xspace}
\newcommand{\TOPlocMRDten}{\ensuremath{5.06^{+1.61}_{-1.31}}\xspace}

\begin{document}

\title{Constraining Population III stellar demographics with next-generation gravitational-wave observatories
}

\author{Cailin Plunkett\,\orcidlink{0000-0002-1144-6708}}
\email{caplunk@mit.edu}

\author{Matthew Mould\,\orcidlink{0000-0001-5460-2910}}
\LIGO\kavli\MIT

\author{Salvatore Vitale\,\orcidlink{0000-0003-2700-0767}}
\LIGO\kavli\MIT

\begin{abstract}
Next-generation gravitational-wave observatories will reach farther into the universe than currently possible, revealing black-hole mergers from early stellar binary systems such as Population III stars, whose properties are currently poorly constrained. We develop a method to infer the properties of their progenitor populations from gravitational-wave catalogs. Using Bayesian deep learning, we train an emulator for population-synthesis predictions of black-hole merger properties across redshift as a function of the initial stellar mass function, crucially accounting for systematic uncertainty due to the finite number of training simulations. Combined with a nonparametric model for star formation history, we analyze catalogs containing both Population I/II and III sources simulated with full Bayesian parameter estimation for a detector network of Cosmic Explorer and Einstein Telescope with one year of observing time. We demonstrate our ability to separate these two populations at high redshifts where both make comparable contributions to the black-hole merger rate, excluding a Population III merger rate of zero at nearly 100\% credibility. Moreover, we can place meaningful constraints on the Population III progenitor distributions; in particular, we constrain the spectral index of the initial mass function to within roughly $\pm0.5$ of the true value and the log of the star formation rate density to within $\sim25\%$ over redshifts 10 to 20. By leveraging astrophysics-informed and astrophysics-agnostic models, we demonstrate the discriminative power of our combined inference approach and highlight the potential of next-generation gravitational-wave observatories to uncover the details of high-redshift stellar populations.
\end{abstract}

\maketitle

\section{Introduction}

Next-generation (XG) gravitational-wave (GW) observatories---namely Cosmic Explorer (CE)~\cite{Reitze:2019iox, Kalogera:2021bya} and the Einstein Telescope (ET)~\cite{Punturo:2010zz, Abac:2025saz}---may detect stellar-mass black-hole (BH) mergers up to redshifts of $z\sim 100$~\cite{Hall2019,Maggiore2020,Branchesi:2023mws,Kalogera:2021bya}.
They are therefore capable of probing BHs that originated as Population III (Pop~III) stars: the first generation of stars that may have formed in the low-metallicity environment of the early universe \cite{Schaerer:2001jc, Glover:2012gx,Bromm:2013iya,Haemmerle:2020iqg,Klessen:2023qmc}.
Pop~III stars are a hypothesized outcome of cosmic dawn, the transitional period after recombination during which the first complex large-scale structures formed. These first stars may have driven reionization, influenced the chemical evolution of the intergalactic medium and provided the chemical enrichment of subsequent stellar populations,
established galactic feedback structures,
and formed the light seeds of supermassive BHs~\cite{Wyithe:2003rr, Sibony:2022tph, Karlsson:2011xx, Dayal:2018hft, Inayoshi:2019fun, Woods:2018lty, Klessen:2023qmc}. However, they are yet to be observed and, because they are dim and short-lived, are likely unresolvable with current or planned telescopes without gravitational lensing~\cite{Rydberg:2012ez,Larkin:2022asx, Schauer:2022ucs,Trussler2023, Venditti2024, Chowdhury:2024wvm}.

Theoretical models disagree on key characteristics of Pop~III stars, including their star formation rate density (SFRD) and initial mass function (IMF). A key unknown for both the SFRD and IMF is the critical mass for the collapse of dark matter halos, which determines when star formation first occurs.
Estimates for the shape and scale of the Pop~III SFRD vary widely based on uncertainties in cosmological and physical parameters, such as the strength of Lyman--Werner background flux, baryon streaming velocities, the gas cooling rate, and multiplicity due to disk fragmentation \cite{Ishiyama:2025tfi, Sugimura:2023knx, Klessen:2023qmc, Schauer:2020gvx}.
Although the rate uncertainty spans three orders of magnitude across numerical simulations~\cite{Jaacks2019,Liu2020,Skinner2020} and semi-analytical models~\cite{Trenti:2009cj,Visbal:2020gri,Hartwig:2022lon,Trinca2024, Ishiyama:2025tfi}, studies consistently find Pop~III star formation begins around redshift $z\sim 30$ and peaks between $z\sim$~10--20, before decreasing as Pop~II star formation begins to dominate~\cite{Klessen:2023qmc}.
Many theoretical models point to a top-heavy initial stellar mass distribution compared to the Pop~I/II stars we observe today, which are enriched with elements heavier than hydrogen and helium \cite{Clark:2010bs,Stacy:2012iz,Susa:2014moa,Stacy2016,Jaura:2022sny}; however, recent constraints from indirect observables suggest the IMF may have a steeper slope than previously thought~\cite{Hartwig:2024wnk}.

Pop~III stars may produce the first BHs born from stellar collapse in the universe and contribute to the population of GW sources across cosmic time; while Pop~III mergers may be a small fraction of the total low-redshift ($z\approx 0$) merger rate, a Pop~III origin has been employed to interpret both high-mass and high-mass-ratio GW events~\cite{Belczynski:2004gu, Kinugawa:2014zha, Hartwig:2016nde, Belczynski:2016ieo, Kinugawa:2020xws, Kinugawa:2020tbg, Iwaya:2023mse}.
BHs that are the remnants of Pop~III stars may be more massive than their Pop~I/II counterparts not only due to their top-heavy IMF, but also since their stellar binary progenitors are more likely to undergo stable mass transfer without experiencing common-envelope episodes \cite{Inayoshi:2017mrs} and low-metallicity stars lose less mass via stellar winds \cite{Madau:2001sc, Volpato:2022nwe}. Due to their high masses, Pop III stars are thought to be candidates for pair-instability supernovae~\cite{Tanikawa:2021qqi, Venditti2024, Tanikawa:2024mpj}. Pop~III stars may also contribute to  astrophysical GW backgrounds~\cite{Kowalska:2012ba,Martinovic:2021fzj}, but their contribution may be minimal, particularly for XG detectors that will resolve most binaries individually~\cite{Kouvatsos:2024eok}.

Since the GW strain from binary BH (BBH) mergers encodes the parameters of the BHs, which in turn encode those of their stellar progenitors, data from XG observatories thus provide an avenue to constrain Pop~III stellar properties.
Such constraints require a mapping between Pop~III stellar distributions and those of their BH remnants. Binary population synthesis (BPS) simulations have been used to explore BBH mergers from Pop~III stars under various assumptions about their progenitor characteristics.
For example, \citet{Costa:2023xsz} explored the range of uncertainties in Pop~III BBH properties by comparing BPS simulations under a variety of assumptions about their initial stellar conditions.
They find the mass distributions of Pop~III BBHs can be similar to those from Pop~II stars, implying it may be hard to distinguish these populations based on mass information alone.
\citet{Santoliquido2023} extended this work by convolving each simulation with several theoretical star formation histories to obtain the redshift-dependent Pop~III BBH distributions. Across binary and star formation uncertainties, they estimate the detection rate with ET to range from 10--$10^4$ mergers per year.

Detecting high-redshift events does not necessarily imply an accurate estimation of their parameters or origin \cite{Ng:2021PBH}. Studies have shown XG observatories are unlikely to provide precise ($\lesssim10\%$) redshift measurements at redshifts $z\gtrsim10$, with worsening measurements at redshifts $z\gtrsim30$ \cite{Vitale:2016icu,Ng:2021PBH,Ng:2022vbz,Mancarella:2023ehn,Santoliquido:2024oqs}. However, \citet{Ng:2020PopIII} demonstrated that Pop~III BBH can be distinguished from other subpopulations in XG catalogs solely on the basis of redshift.
Additionally, \citet{Santoliquido:2024oqs} presented a machine-learning classifier for Pop~III and Pop~I/II BBHs with approximated parameter-estimation uncertainties and found a nonzero fraction of Pop~III events can be correctly identified, with the proportion depending on the Pop~III population parameters. Combining information across multiple parameters is likely the most fruitful approach for obtaining reliable constraints on high-redshift GW sources.

Inferring population-level characteristics from GW data typically involves Bayesian hierarchical inference with an assumed model for the underlying population. Broadly speaking, GW population models lie on a spectrum from ``astrophysics-agnostic'' to ``astrophysics-informed.'' On the agnostic side, nonparametric models seek to avoid strong prior assumptions about the population and thus describe the data with minimal bias \cite[e.g.,][]{Heinzel:2023hlb, Edelman:2022ydv, Golomb:2022bon, Godfrey:2023oxb, Rinaldi:2021bhm, Mandel:2016prl, Farah:2024xub, Ray:2023upk, Tiwari:2020vym, Toubiana:2023egi, Callister:2023tgi, Heinzel:2024jlc, Sadiq:2023zee}. These flexible models are the most agnostic about the underlying astrophysics but often suffer from a lack of interpretability, higher measurement uncertainty, and increased computational expense.
In the middle, parametric models specify simple functional forms to describe the population based on expected features. For instance, a commonly used population model for BBH spin directions is a mixture between isotropic and aligned components representing dynamical and isolated binary formation, respectively~\cite{Vitale:2015tea, Stevenson:2017dlk, Talbot:2017yur}.
The common \textsc{Power Law + Peak} mass model consists of a power law underlying a Gaussian that is intended to capture a possible pileup from pulsational pair-instability supernovae~\cite{Talbot:2018cva}.
While parametric models are simple to implement, they do not directly relate to the underlying physics and, further, can only describe the data as flexibly as the function allows, with potential for mis-specification \cite{Romero-Shaw:2022ctb}.
At the other end of the spectrum, simulation-based methods use the inputs of astrophysical simulations as parameters of the population model, allowing for direct inference on the physics of binary evolution or features of stellar populations from GW catalogs \cite[e.g.,][]{Barrett:2016edh, Taylor:2018iat, Wong:2020ise, Mould:2022ccw, Riley:2023jep, Leyde:2023iof, Mastrogiovanni:2022ykr, Zevin:2020gbd, Colloms:2025hib}.
However, of the three approaches, astrophysics-informed analyses make the strongest assumptions about the underlying population, which can introduce biases if false~\cite{Cheng:2023ddt}. 
Nonetheless, they offer otherwise unobtainable insights into which astrophysical parameters and populations may best describe the GW data.

In this work, we assess the capabilities of XG detectors with regard to distinguishing and measuring population-level features of Pop~I/II and III BBHs. We develop a method to infer the SFRD and IMF of Pop~III stars directly from the catalog of detected GW events, linking nonparametric, parametric, and astrophysical models to leverage each of their strengths while encoding varying levels of prior assumptions. Using full Bayesian parameter estimation with a CE--ET detector network, we run hierarchical inference on simulated catalogs spanning one year of observing time that contain both Pop~I/II and Pop~III BBH sources and demonstrate our ability to simultaneously constrain the properties of both populations.

This paper is organized as follows.
In Section~\ref{sec:astropop} we introduce the assumed astrophysical populations of Pop~I/II and III stars that make up our simulated universes.
We describe the details of our mixed population modeling approach in Section~\ref{sec:popmodel} and present the results on simulated XG GW catalogs in Section~\ref{sec:results}.
Finally, we conclude and point to future applications of this work in Section~\ref{sec:conclusion}.

\section{Astrophysical populations}
\label{sec:astropop}

As the tail end of Pop~III star formation is thought to coincide with the onset of Pop~I/II star formation, we expect high-redshift ($z\gtrsim10$) XG sources to comprise at least two subpopulations: Pop~I/II and Pop~III. As such, we construct simulated universes containing mergers from both.
For the BBH source parameters not discussed below, we assume their underlying population follows typical uninformative distributions, e.g., isotropic sky location.

\subsection{Pop~I/II binary black holes}
\label{sec:localstars}

The local MRD inferred from current LIGO--Virgo--KAGRA (LVK) data \cite{GWTC3, GWTC3Pop} is much higher than that predicted for local Pop~III BH mergers \cite{Santoliquido2023}. We therefore assume that the currently constrained population consists only of Pop~I/II mergers and select the corresponding properties of our simulated population to match current constraints.
Although the LVK population may contain Pop~III mergers with long time delays, they are likely a small fraction of observed mergers~\cite{Belczynski:2016ieo}, especially as current constraints indicate a preference for short time delays~\cite{Fishbach:2021mhp, Schiebelbein-Zwack:2024roj, Karathanasis:2022rtr}, although these results depend on assumptions for the star formation rate of the population.
Our Pop~I/II model and its parameters are summarized in Table~\ref{tab:PopIProps} and described below.

\textbf{Pop~I/II masses:}
We assume the underlying mass distribution for Pop~I/II source follows the \textsc{Power Law + Peak} model introduced in Ref.~\cite{Talbot:2018cva}: the distribution of primary masses $m_1$ (i.e., the heavier BH in the binary) is a mixture between a doubly truncated power law and a Gaussian, with a smooth turnon at low masses; the mass ratio ($0<q\leq1$) distribution is a power law such that the secondary mass $m_2\leq m_1$ has the same minimum BH mass and smoothing as $m_1$. We use parameters consistent with the results from the third gravitational-wave transient catalog (GWTC-3)~\cite{GWTC3Pop}; see Table~\ref{tab:PopIProps}.

\textbf{Pop~I/II redshifts}:
The BBH MRD is poorly constrained at high redshifts due to the limited sensitivity of the LVK detectors above $z\sim 1$. We adopt the estimates of~\citet{Callister:2020arv} as our Pop~I/II BH merger rate. Assuming the comoving MRD follows the rate of stellar formation with a log-uniform time-delay distribution, the predicted BBH merger rate is well fit by a Madau--Dickinson model~\cite{Madau:2014bja}:
\begin{align}\label{eq:MD}
\mathcal{R}(z)
=
\mathcal{R}_0
\left[ 1+(1+z_p)^{-\lambda-\kappa} \right]
\frac
{ (1+z)^\lambda }
{ 1 + \left( \frac{1+z}{1+z_p} \right)^{\lambda+\kappa} }
\, ,
\end{align}
where $\lambda = 1.9$ is the low-redshift power-law index $\mathcal{R}(z)\propto(1+z)^\lambda$ and $\kappa = 3.4$ is the high-redshift power-law index $\mathcal{R}(z)\propto(1+z)^{-\kappa}$ that turns over around $z_p = 2.4$. The normalization is such that $\mathcal{R}(0)=\mathcal{R}_0$. We select $\mathcal{R}_0 = 19~\mathrm{Gpc}^{-3}\mathrm{yr}^{-1}$, consistent with the inferred rate in GWTC-3~\cite{GWTC3Pop}. This MRD is shown in orange in the bottom panel of Figure~\ref{fig:true_sfrds_mrds}.

\textbf{Pop~I/II spins:}
For the spin magnitudes $a_1$ and $a_2$, we assume independent and identically distributed (IID) draws from a normal distribution truncated between $0 \leq a_{1,2} < 1$ with location and scale parameters $\mu_a=0$ and $\sigma_a=0.3$, respectively.
For the polar angles $\tau_{1,2}$ between the BH spin vectors and the orbital angular momentum, we assume preferentially aligned IID truncated Gaussians for $-1 \leq \cos\tau_{1,2} \leq 1$ with location parameter $\mu_\tau=1$ and scale $\sigma_\tau=1.2$.
Though parametrized by different models, the values of these parameters are chosen to produce spin distributions consistent with those inferred in Ref.~\cite{GWTC3Pop}.

\begin{table}
\centering
\begin{tabular}{ccc} \hline \hline 
\textbf{Parameter} & \textbf{Description} & \textbf{Truth} \\ \hline
$\alpha_m$ & $m_1$ spectral index & 3.4 \\
$\beta_q$ & $q$ spectral index & 1.1 \\
$\lambda_m$ & Peak fraction & 0.04 \\
$\mu_m$ & Peak location [$M_\odot$] & 34 \\
$\sigma_m$ & Peak width [$M_\odot$] & 3.6 \\
$m_\mathrm{min}$ & Minimum BH mass [$M_\odot$] & 5.0 \\
$m_\mathrm{max}$ & Maximum BH mass [$M_\odot$] & 87 \\
$\delta_m$ & Low-mass smoothing [$M_\odot$] & 4.8 \\ \hline
$\mathcal R_0$ & Local rate [Gpc$^{-3}$yr$^{-1}$] & 19 \\
$\lambda$ & Low-$z$ spectral index & 1.9 \\
$\kappa$ & High-$z$ spectral index & 3.4 \\
$z_p$ & Redshift peak & 2.4 \\ \hline
$\mu_a$ & Spin magnitude peak & 0 \\
$\sigma_a$ & Spin magnitude width & 0.3 \\
$\mu_\tau$ & Cosine tilt peak & 1 \\
$\sigma_\tau$ & Cosine tilt width & 1.2 \\\hline \hline
\end{tabular}
\caption{Parameters for the mass (top), redshift (middle), and spin (bottom) distributions for our simulated population of Pop~I/II sources.}
\label{tab:PopIProps}
\end{table}

\subsection{Pop~III stars}
\label{sec:popIIIstars}

\textbf{Pop~III masses:}
We employ the Pop~III simulations of~\citet{Costa:2023xsz}, who used \textsc{SEVN} \cite{Iorio:2022sgz, Spera:2017fyx} to explore the effect of the stellar IMF and initial distributions of mass ratio, orbital period, and eccentricity at metallicity $Z=10^{-11}$ on the resulting population of merging BBHs.
We consider IMFs for zero-age main sequence (ZAMS) masses $M$ from the two-parameter family
\begin{equation}
\label{eq:imf}
p \left( \frac{M}{M_\odot} \right)
\propto
\left( \frac{M}{M_\odot} \right)^{-\alpha}
\exp \left[ -20\left(\frac{M}{M_\odot} \right)^{-\beta} \right]
\, ,
\end{equation} where here and throughout $p$ is used to denoted a probability density function (PDF).
We simulate distinct catalogs for two choices of this IMF, which are shown in boldface in Table~\ref{tab:IMFparams}:
\begin{itemize}
\item a top-heavy IMF (TOP) with $\alpha=0.17$ and $\beta=2$;
\item a log-uniform IMF (LOG) with $\alpha=1$ and $\beta=0$.
\end{itemize}
In both cases, fits to O and B stars in the local universe by \citet{Sana:2012px} are used for the initial stellar binary distributions of:
\begin{itemize}
\item mass ratios $Q$\footnote{We use $Q$ for stellar mass ratio to differentiate it from BBH mass ratio $q$.}$\in[0.1,1]$, where $p(Q) \propto Q^{-0.1}$;
\item orbital periods $P$ given by $\Pi := \log(P/\mathrm{day}) \in [0.15, 5.5]$ with $p(\Pi) \propto \Pi^{-0.55}$;
\item eccentricities $e\in[0,1)$ with $p(e) \propto e^{-0.42}$.
\end{itemize}
Each simulation results in a set of BH masses and time delays of merging BBHs from the Pop~III progenitors.

\begin{table}
\centering
\setlength{\tabcolsep}{6pt}
\begin{tabular}{ccc}
\hline\hline
Model & $\alpha$ & $\beta$ \\
\hline
\textbf{LOG} & \textbf{1} & \textbf{0} \\
\textbf{TOP} & \textbf{0.17} & \textbf{2} \\
KRO & 2.3 & 0 \\
LAR & 2.35 & 1 \\
\hline\hline 
\end{tabular}
\caption{Parameters of the ZAMS IMF from Eq.~(\ref{eq:imf}) used  in Refs.~\cite{Costa:2023xsz, Santoliquido2023} for Pop~III simulations. From top to bottom: flat-in-log masses (LOG), top-heavy IMF (TOP), a Kroupa mass function \cite{Kroupa:2000iv} (KRO), and Larson distribution \cite{Larson:1998iu} (LAR). We highlight in bold the models we use for our two simulated universes. All four simulations are used to train a surrogate population model for hierarchical inference, as described in Sec.~\ref{sec:popmodel}.}
\label{tab:IMFparams}
\end{table}

\textbf{Pop~III redshifts:}
\citet{Santoliquido2023} use \textsc{CosmoRate}~\cite{Santoliquido:2020axb} to compute the MRDs (see Fig.~3 of Ref.~\cite{Santoliquido2023}) corresponding to each Pop~III simulation in Ref.~\cite{Costa:2023xsz} by convolving their time-delay distributions with different theoretical SFRDs (see Fig.~1 of Ref.~\cite{Santoliquido2023}). They note that this introduces an inconsistency because each of those SFRDs assume IMFs that may differ from those used in the BPS simulations. To ensure consistency, we select the SFRDs---and corresponding MRDs and binary properties---that use the same IMFs as the BPS simulations.

\begin{figure}
\centering
\includegraphics[width=0.9\columnwidth]{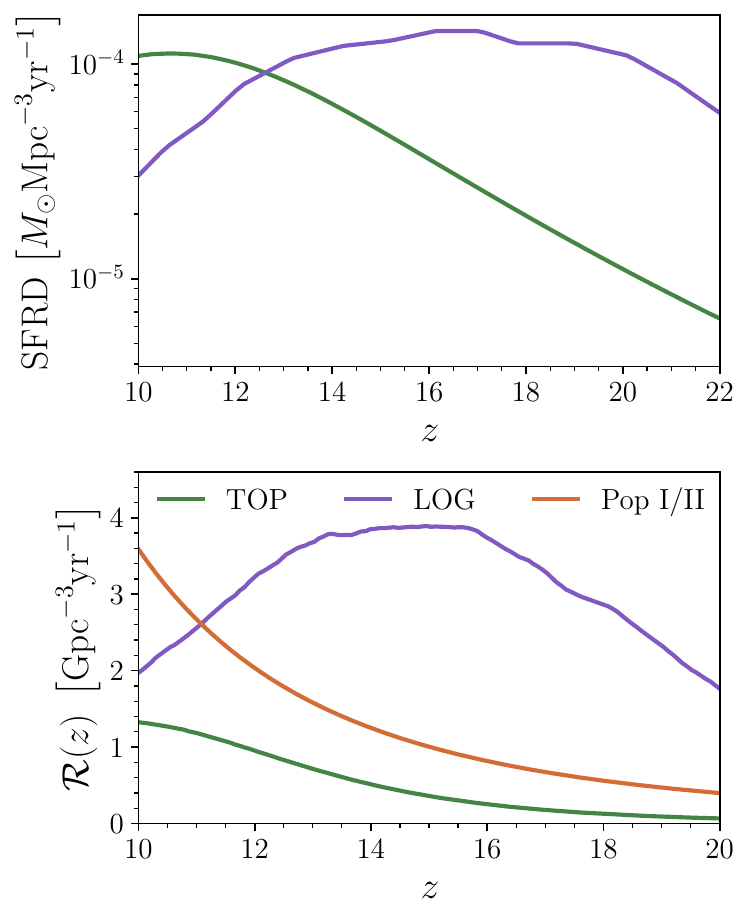}
\caption{The SFRDs (top panel) and MRDs (bottom panel) for the TOP (\TOPclr) and LOG (\LOGclr) Pop~III models; the local, Pop I/II MRD is also shown in orange. The SFRDs are derived by~\citet{Liu2020} and ~\citet{Hartwig:2022lon} for TOP and LOG, respectively, and the MRDs are computed by~\citet{Santoliquido2023}. The local MRD uses the functional form of~\citet{Callister:2020arv} with scale from~\citet{GWTC3Pop}.}
\label{fig:true_sfrds_mrds}
\end{figure}

For the TOP IMF we use the Pop~III SFRD derived by \citet{Liu2020} who ran a cosmological simulation with \textsc{Gizmo}~\cite{Hopkins:2014qka} and fit their results to a Madau--Dickinson function~\cite{Madau:2014bja}.
For LOG we use the results of \citet{Hartwig:2022lon} who use the semi-analytical \textsc{A-Sloth} code. We refer to these SFRDs also by the IMFs they use: TOP and LOG. We plot these SFRDs and resulting MRDs in Figure~\ref{fig:true_sfrds_mrds}. The LOG SFRD dominates at high redshifts, peaking at $z\approx16.5$ and reaching up to an order of magnitude above the TOP SFRD, which peaks at $z\approx10.5$. In addition, the LOG IMF model has a merger efficiency a factor of $\sim3$ larger than the TOP model. Coupled with its higher SFRD, the merger rate density of LOG is larger than that of TOP for $z>9$ and is $\approx25$ times higher at $z=20$.

\textbf{Pop~III spins:}
The Pop~III BPS simulations by \citet{Costa:2023xsz} and \citet{Santoliquido2023} do not include stellar or BH spins. However, Pop~III stars may be rapidly rotating~\cite{Stacy:2012iz,Yoon2012}, which would contribute to nonzero BH spins. So, we adopt the same spin distributions as for Pop~I/II BBHs above.
This is a conservative choice in the sense that it makes the two populations more difficult to distinguish.

\subsection{Pop~III surrogate}
\label{sec:surrogates}

\begin{figure*}
\centering
\includegraphics[width=0.9\columnwidth]{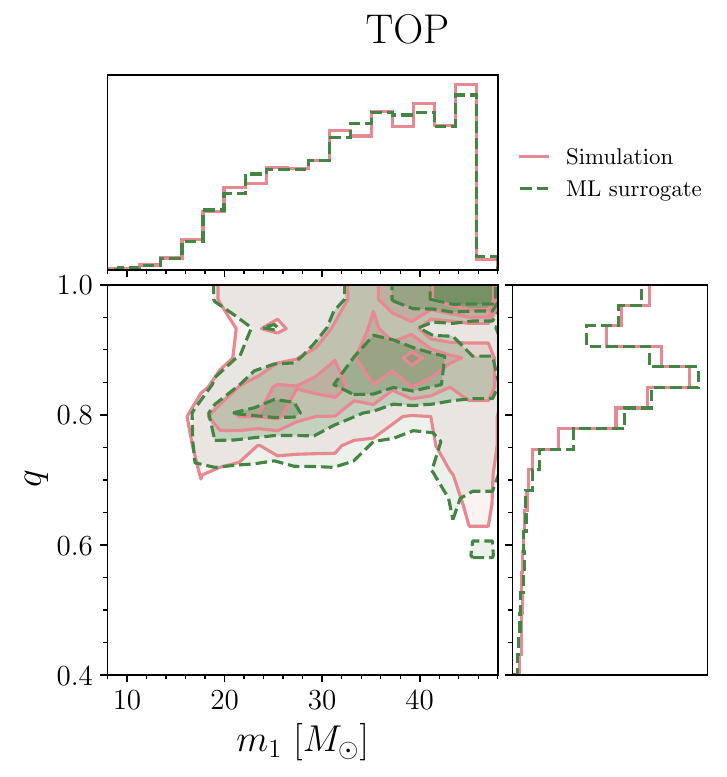}
\hspace{1cm}
\includegraphics[width=0.9\columnwidth]{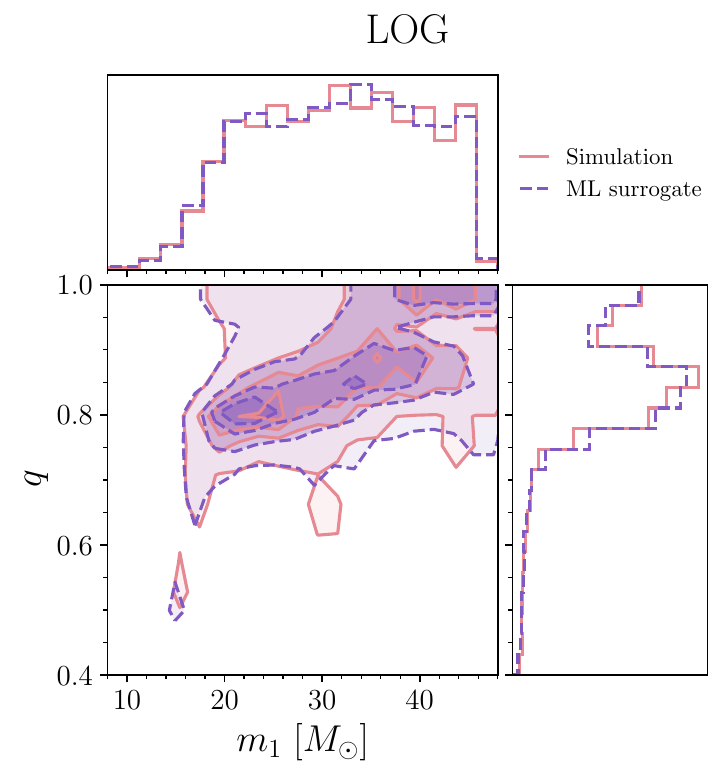}
\caption{Joint two-dimensional distributions (central panels) and one-dimensional marginalizations (upper- and right-hand panels) of primary BH mass $m_1$ and BBH mass ratio $q$ for the TOP (left) and LOG (right) Pop~III models at redshift $z=10$. The population synthesis simulation data used for training are shown in pink. Samples from the normalizing flow emulators are in dashed green and purple for TOP and LOG, respectively. For the joint $m_1$--$q$ distributions we show the $(0.5, 1, 1.5, 2)$-$\sigma$ equivalent levels, corresponding to the 11.8\%, 39.3\%, 67.5\% and 86.4\% credible regions.
}
\label{fig:simulations}
\end{figure*}

Including the two Pop~III simulations discussed above, \citet{Santoliquido2023} considered a total of four IMFs corresponding to different values of $\alpha$ and $\beta$, as listed in Table~\ref{tab:IMFparams}.
For each, they published catalogs of the masses and time delays of BBH mergers at 18 discrete redshift values, each containing $\sim2\times10^5$ mergers. To generate sources continuously across redshift for our GW catalogs, we fit a surrogate model to the Pop~III simulations with normalizing flows, which use neural networks to parameterize (conditional) probability distributions \cite{JMLR:v22:19-1028, 9089305}. In particular, we train the flow to emulate the distribution $p(m_1,q|z,\alpha,\beta)$, allowing us to evaluate the PDF and draw new $m_1$ and $q$ samples continuously over different redshifts $z$ and IMFs determined by $\alpha$ and $\beta$. In principle, this distribution also depends on the assumed SFRD; however, we found it is primarily controlled by the mass-dependent time-delay distribution induced by the IMF and that neglecting dependence on the SFRD had negligible impact.

Machine-learning models carry inherent systematic uncertainty from the finite number of input samples and stochastic training. Indeed, we found that flows trained with different initializations predict slightly different mass distributions $p(m_1,q|z,\alpha,\beta)$ for $\alpha$ and $\beta$ values in the training set and notably dissimilar results for other values. To better calibrate the flow predictions at the TOP and LOG IMFs we consider, we use an ensemble training approach~\cite{NIPS2017_9ef2ed4b} described in detail in Section~\ref{sec:p3mm}. In short, we trained $W=50$ separate normalizing flows on the same training dataset but using different random initializations. Denoting the parameters (neural-network weights) of each flow as $w_i$ ($i=1,...,W$), we take the ensemble average
\begin{align}
p(m_1,q|z,\alpha,\beta)
=
\frac{1}{W}
\sum_{i=1}^W p(m_1,q|z,\alpha,\beta,\omega_i)
\,
\label{eq: ensemble average}
\end{align} for the astrophysical populations in place of the BPS simulations. Averaging over this ensemble of flows produces better-calibrated predictions not only for the TOP and LOG IMFs we consider---which lie within the training set---but for other values across the IMF parameter space, which is key for population inference. Further details on normalizing flows are provided in Appendix~\ref{app:flows}.

We compare the machine-learning (ML) predictions to the two Pop~III simulations we use for our astrophysical populations in Figure~\ref{fig:simulations}, showing the primary mass and mass ratio BBH distributions at $z=10$.
The surrogates purposefully smooths some of the features present in the simulation catalogs that occur due to finite sampling.
Both the TOP and LOG populations contain clusters at $q\approx0.8$ from binaries that underwent stable mass transfer with no common-envelope episodes and at nearly equal masses corresponding to systems that experienced at least one common envelope after a stable mass transfer episode.

\subsection{Simulated catalogs}
\label{sec:catalogs}

We create catalogs corresponding to $T=1$~yr of observation time with an XG detector network of a triangular ET detector and 40~km and 20~km CE detectors with currently projected sensitivities and standard locations \cite{T2300003}.
Signals were generated using the \textsc{IMRPhenomXPHM} waveform model~\cite{Pratten:2020ceb} and selected via a detection threshold of a network matched-filter signal-to-noise ratio (SNR) of 11. As the Pop~I/II merger rate dominates below $z\lesssim10$, we would need to consider a prohibitively large total number of detections to generate any significant number of Pop~III BBH mergers. We therefore only consider events in the redshift range $z\in[10,20]$, where each population makes a comparable contribution to the overall rate. If we were to increase the redshift range, we would gain constraining power on the Pop~I/II binaries, but likely add minimal information on the Pop~III objects of interest.
Using the MRDs from the two distinct models we consider, this corresponds to GW catalogs with 138 Pop~I/II events, one with 77 Pop~III events from the TOP model and another with 350 Pop~III events from the LOG model.

For each event, we perform full Bayesian parameter estimation with \textsc{bilby}~\cite{bilby_paper} using the accelerated relative binning gravitational-wave likelihood~\cite{relbin_bilby, relbin_cornish, relbin_zackay} and the \textsc{dynesty} nested sampler~\cite{Speagle:2019ivv}. We use uniform priors in detector-frame total mass, mass ratio, spin magnitudes and cosine tilts, as well as standard priors for the extrinsic parameters~\cite{Romero-Shaw:2020owr}.

\section{Population inference}
\label{sec:popmodel}

We analyze our simulated catalogs using Bayesian hierarchical inference \cite{Mandel:2018mve, Vitale2020}, the details of which are presented in Appendix~\ref{app:likelihood}. Whereas in the previous section we described the models for the ``true'' population used to generate mock GW catalogs, here we describe the population models we use for inference.

The key ingredient in GW population analyses is the differential detector-frame merger rate $R(\theta,z|\Lambda) = \dd{N} / (\dd{\theta} \dd{z} \dd{t_d})$ as a function of the BBH parameters $\theta$, redshift $z$, and population parameters $\Lambda$, where $N$ is the number of BH mergers and $t_d$ represents detector-frame time. This is related to the comoving MRD $\mathcal{R}(\theta,z|\Lambda) = \dd{N} / (\dd{\theta} \dd{V_c} \dd{t_s})$ that we model directly by
\begin{align}\label{eq:MRD}
R(\theta,z|\Lambda)
=
\mathcal{R}(\theta,z|\Lambda)
\frac {\dd{V_c}}{\dd{z}}
\frac{1}{1+z}
\, ,
\end{align}
where $V_c$ is the comoving volume and $t_s$ the source-frame time. Using the GW catalog data denoted $\mathcal{D}$, the goal of population inference is to infer the parameters $\Lambda$ by using Bayes' theorem to find the posterior distribution $p(\Lambda|\mathcal{D}) \propto p(\mathcal{D}|\Lambda) p(\Lambda)$, where $p(\mathcal{D}|\Lambda)$ is the likelihood and $p(\Lambda)$ the prior.

Our catalogs contain sources from both the Pop~I/II and Pop~III populations, hence we fit the simulated data using a model that captures both subpopulations:
\begin{align}
\mathcal{R}(\theta,z|\Lambda)
=
\mathcal{R}_\mathrm{I/II}(\theta,z|\Lambda)
+
\mathcal{R}_\mathrm{III}(\theta,z|\Lambda)
\, .
\end{align}
We further factor both components of the model into a redshift-dependent MRD and PDFs to describe the shape of the population over masses and spins. We also assume the BH masses and spins are conditionally independent and that spins are IID and do not evolve over redshift; i.e.,
\begin{align}
\mathcal{R}_\mathrm{X}(\theta,z|\Lambda)
& =
\mathcal{R}_\mathrm{X}(z|\Lambda)
p_\mathrm{X}(m_1,q|z,\Lambda)
\nonumber
\\
& \times
p_\mathrm{X}(a_1|\Lambda)
p_\mathrm{X}(a_2|\Lambda)
p_\mathrm{X}(\tau_1|\Lambda)
p_\mathrm{X}(\tau_2|\Lambda)
\,,
\end{align}
where the subscript X denotes the subpopulation, I/II or III.

\subsection{Pop~I/II BBH model}
\label{sec:locmodels}

The primary target of this study is the Pop~III subpopulation. For the Pop~I/II BBH, an increasing number of detections with current GW detectors \cite{Baibhav:2019gxm, Broekgaarden:2023rta} will lead to improved constraints on the low-redshift population of sources before the next-generation era. Assuming the inferred redshift evolution can be extrapolated back to the range $z\in[10,20]$ that we consider, we will develop a good handle on the Pop~I/II BBH population. Therefore, we use similar parametric models for inference as the underlying Pop~I/II distributions described in Sec.~\ref{sec:localstars}, the important difference being that we fit for the parameters in Table~\ref{tab:PopIProps} rather than use their fixed values.
In particular, $p_\mathrm{I/II}(m_1,q|z,\Lambda) = p_\mathrm{I/II}(m_1,q|\Lambda)$ is the \textsc{Power Law + Peak} model, $p_\mathrm{I/II}(a_{1,2}|\Lambda)$ is a truncated Gaussian for $0 \leq a_{1,2} < 1$, and $p_\mathrm{I/II}(\tau_{1,2}|\Lambda)$ is a truncated Gaussian for $-1 \leq \cos\tau_{1,2} \leq 1$.
Because we anticipate the Pop~I/II merger rate to peak well below $z=10$, rather than use the full four-parameter Madau-Dickinson model in Eq.~(\ref{eq:MD}),
we reduce to its high-redshift behavior, which follows a power-law
\begin{align}
    \mathcal{R}_\mathrm{I/II}(z|\Lambda) = \mathcal R_{10} \frac{(1+z_0)^\kappa}{(1+z)^\kappa},
\end{align}
which we have normalized to $z=10$ by setting $\mathcal R_{10}=\mathcal R_\mathrm{I/II}(10)$ and $z_0=10$.
The priors we use are listed in Table~\ref{tab:priors}.

\begin{table}
\begin{tabular}{cccc} \hline \hline
\textbf{Parameter} & \textbf{Pop.} & \textbf{Description} & \textbf{Prior} \\ \hline \hline
$\alpha_m$ & I/II & $m_1$ spectral index & $\mathcal U(-1, 8)$ \\
$\beta_q$ & I/II & $q$ spectral index & $\mathcal U(-1, 4)$ \\
$\lambda_m$ & I/II & Peak fraction & $\mathcal U(0,1)$\\
$\mu_m$ & I/II & Peak location [$M_\odot$] & $\mathcal U(20, 50)$ \\
$\sigma_m$ & I/II & Peak width [$M_\odot$] & $\mathcal U(1,8)$\\
$m_\mathrm{min}$ & I/II & Minimum BH mass [$M_\odot$] & $\mathcal U(2,10)$\\
$m_\mathrm{max}$ & I/II & Maximum BH mass [$M_\odot$] & $\mathcal U(50,100)$\\
$\delta_m$ & I/II & Low-mass smoothing [$M_\odot$]& $\mathcal U(0,8)$\\ \hline
$\mathcal R_{10}$ & I/II & MRD at $z=10$ [Gpc$^{-3}$yr$^{-1}$] & $\mathcal U(1,8)$ \\
$\kappa$ & I/II & Redshift spectral index & $\mathcal U(1,10)$ \\

\hline
$\mu_\mathrm{I/II}^a$ & I/II & Spin magnitude peak & $\mathcal U(0,0.3)$ \\
$\sigma_\mathrm{I/II}^a$ & I/II & Spin magnitude width & $\mathcal U(0.1,0.5)$ \\
$\mu_\mathrm{I/II}^\tau$ & I/II & Spin tilt peak & $\delta(1)$ \\
$\sigma_\mathrm{I/II}^\tau$ & I/II & Spin tilt width & $\mathcal U(0.5,2.5)$ \\ \hline \hline
$\alpha_\mathrm{IMF}$ & III & IMF spectral index & $\mathcal U(-2,4)$\\
$\beta_\mathrm{IMF}$ & III & IMF exponential taper & $\mathcal U(-1,3)$\\ \hline
$\psi_i$ & III & SFRD at $z_i$ [$M_\odot\,\mathrm{Mpc}^{-3}\,\mathrm{yr}^{-1}]$ & GP \\
\hline
$\mu_\mathrm{III}^a$ & III & Spin magnitude peak & $\mathcal U(0,0.3)$\\
$\sigma_\mathrm{III}^a$ & III & Spin magnitude width & $\mathcal U(0.1,0.5)$\\
$\mu_\mathrm{III}^\tau$ & III & Spin tilt peak & $\delta(1)$ \\
$\sigma_\mathrm{III}^\tau$ & III & Spin tilt width & $\mathcal U(0.5,2.5)$ \\ \hline \hline
\end{tabular}
\caption{Parameters of the models used for Bayesian population inference and their priors. We use $\mathcal{U}$ to denote a uniform distribution, $\delta$ for a delta-function prior, and GP for a Gaussian-process prior (App.~\ref{app:gp}).}
\label{tab:priors}
\end{table}

\subsection{Pop~III BBH model}
\label{sec:p3models}

As seen in Fig.~\ref{fig:simulations}, the complex correlations between Pop~III BBH parameter distributions make them difficult to model with simple functional forms, further motivating the use of more sophisticated population-modeling techniques. For the Pop~III subpopulation, we therefore develop a novel mixed-modeling approach that leverages the advantages of both astrophysics-informed and astrophysics-agnostic models.

Binary stellar evolution does not depend on redshift itself, but on astrophysical quantities that in turn evolve over redshift, such as metallicity. The main uncertainties in BPS modeling are thus the details of single-star evolution and binary processes \cite{Hurley:2002rf, Belczynski:2005mr, Giacobbo:2018etu, Kruckow:2018slo, Spera:2018wnw, Breivik:2019lmt, COMPASTeam:2021tbl, Andrews:2024saw}. Conditioned on modeling assumptions for these uncertain evolutionary processes, BPS simulations provide clear maps from input stellar population properties, such as the IMF, to merging BBH distributions. On the other hand, the history of star formation over cosmic time is probed by observations of star-forming galaxies, only recently reaching to the redshifts $z>10$ \cite{2024Natur.633..318C} targeted in this work. Theoretical predictions for high-redshift star formation history span semi-analytic models to cosmological simulations \cite{Liu2020,Hartwig:2022lon,Jaacks2019,Skinner2020}, which do not take a consistent set of cosmological inputs or physical prescriptions. We therefore opt to use a simulation-based model for the Pop~III BBH mass distribution but an uninformed nonparametric model for the Pop~III SFRD, which in turn is used to compute the MRD. As the BPS simulations do not model BH spins, we use the same spin model as the Pop~I/II distribution but assume independent parameters; see Table~\ref{tab:priors}.

\subsubsection{Pop~III masses: simulation-based ensemble}
\label{sec:p3mm}

To infer the IMF parameters $\alpha,\beta$, we aim to compare the distribution $p(m_1, q | z, \alpha, \beta)$ predicted by BPS to our simulated GW catalog data. However, since BPS is computationally costly, running a simulation for every proposed value of $\alpha,\beta$ would be prohibitively expensive. As we have constructed a Pop~III BPS emulator in Sec.~\ref{sec:surrogates} that depends continuously on $\alpha$ and $\beta$, we can use this model in population inference to constrain the Pop~III IMF.
As before, it is crucial to account for the additional source of uncertainty from training the surrogate model. Intuitively, one would expect the measurement of the population parameters $\Lambda$---in particular, the IMF parameters $\alpha$ and $\beta$---to be more uncertain due to the surrogate uncertainty. However, if we were to use the ensemble average from Eq.~(\ref{eq: ensemble average}) as our population model, the posterior variance would be the same as when using a single flow from the ensemble: averaging calibrates the model in uncertain regions of the IMF parameter space but does not propagate that uncertainty through inference.

Instead, we develop a novel Bayesian approach by modifying the population posterior introduced at the start of Sec.~\ref{sec:popmodel} to explicitly condition on the surrogate training data $\mathcal{T}$, i.e., $p(\Lambda|\mathcal{D},\mathcal{T})$. We consider the neural-network weights $\omega$ as random variables \cite{10.1162/neco.1992.4.3.448, 9756596} in the Bayesian hierarchical model alongside the astrophysical parameters of interest $\Lambda$. Since the GW data have no bearing on the surrogate training, we have $p(\omega|\mathcal{D},\mathcal{T}) = p(\omega|\mathcal{T})$. Similarly, the training data only influence population inference through the network weights, i.e., $p(\Lambda|\mathcal{D},\mathcal{T},\omega) = p(\Lambda|\mathcal{D},\omega)$. Together, these yield
\begin{align}
p(\Lambda|\mathcal{D},\mathcal{T})
=
\int \dd{\omega}
p(\Lambda|\mathcal{D},\omega) p(\omega|\mathcal{T})
\, .
\label{eq: ensemble posterior}
\end{align}
The term $p(\omega|\mathcal{T})$ is the Bayesian posterior distribution of the network weights given the training setup; the corresponding likelihood function is simply given by the flow PDF on the training data. Performing full Bayesian inference to derive this distribution via Hamiltonian Monte Carlo \cite{pmlr-v139-izmailov21a} was not feasible even with weight reparametrization \cite{NIPS2016_ed265bc9, pourzanjani2017improving}, so we instead reuse the ensemble of flows by assuming each $\omega_i$ is a fair draw from the posterior $p(\omega|\mathcal{T})$.
This approximation is not strictly Bayesian but often outperforms truly Bayesian approaches for robust uncertainty quantification \cite{pmlr-v108-pearce20a, NEURIPS2019_8558cb40, NEURIPS2020_322f6246}. Compared to approximations based on the Fisher information \cite{LY201740} or variational inference \cite{Blei03042017} that we found only probed weights local to a single mode of the posterior $p(\omega|\mathcal{T})$, the ensemble covered the range of uncertainty much better.
Finally, we use a coarse-grained Monte Carlo approximation for the integral in Eq.~(\ref{eq: ensemble posterior}):
\begin{align}
p(\Lambda|\mathcal{D},\mathcal{T})
=
\frac{1}{W} \sum_{i=1}^W
p(\Lambda|\mathcal{D},\omega_i)
\, .
\end{align}
In words, we perform separate population-inference runs in which we use a single flow from the ensemble as the Pop~III BBH population model. Using the flow with network parameters $\omega_i$ produces samples from the posterior $p(\Lambda|\mathcal{D},\omega_i)$. We then combine the results of each run with equal weighting to marginalize over the surrogate uncertainty.
In GW data analysis, similar approaches have been used to account for uncertainty in detector noise properties for single-event parameter estimation~\cite{Biscoveanu:2020kat}.
Contrasting against Eq.~(\ref{eq: ensemble average}), we see Eq.~(\ref{eq: ensemble posterior}) is the ensemble-averaged posterior for the surrogates as opposed to the posterior of the ensemble-averaged surrogate. Note that this means none of the recovery models exactly match the true population.

\begin{figure*}[!htb]
\centering
\includegraphics[width=0.9\textwidth]{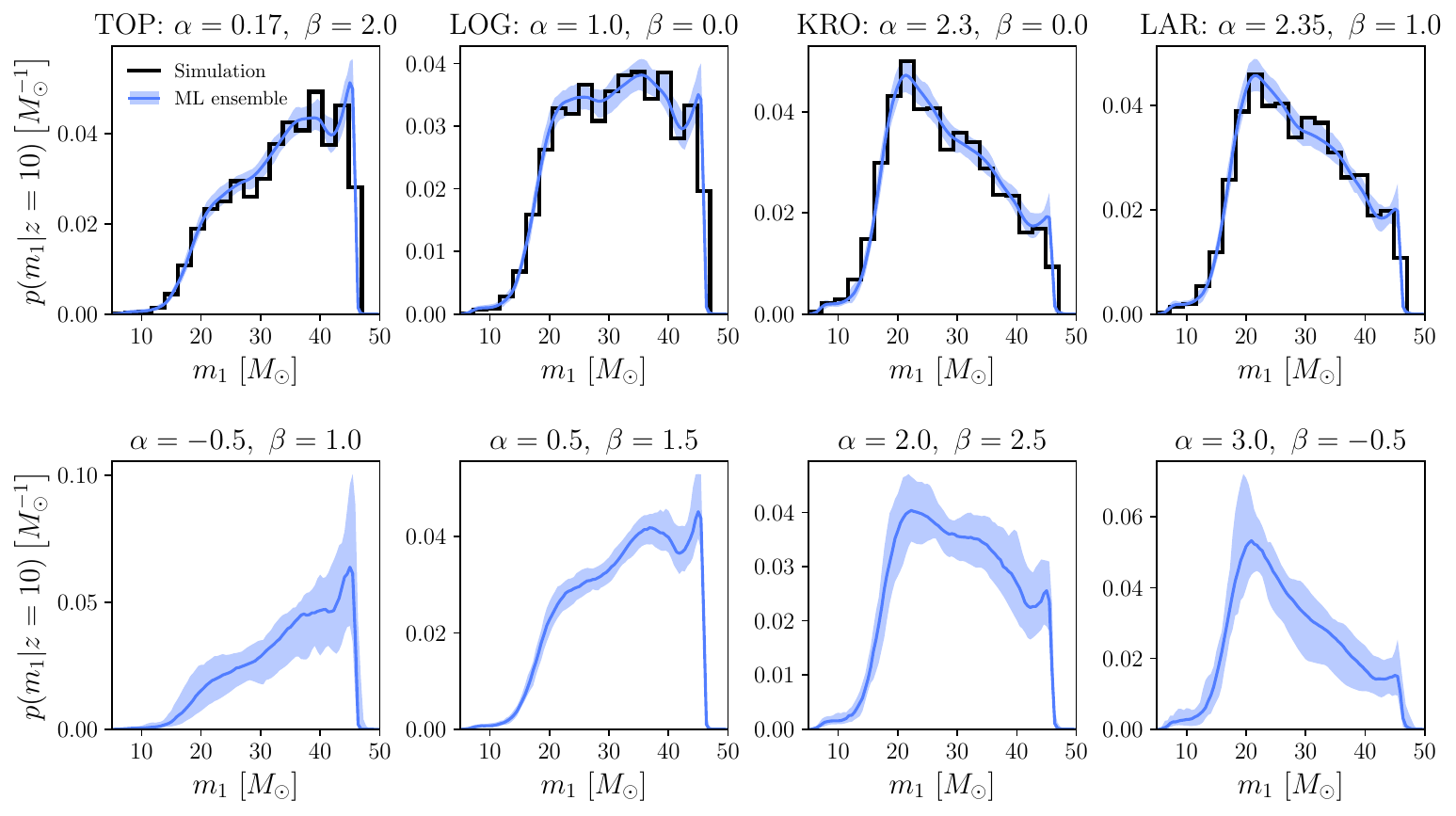}
\caption{Primary mass distributions at $z=10$ for different values of the IMF parameters $(\alpha, \beta)$, as predicted by the ensemble of normalizing flows. The blue lines show the mean of the flow predictions, which we use for the true population (see Section~\ref{sec:surrogates}); the surrounding bands span the 90\% credible regions. The top row shows the predictions for the four values of $(\alpha,\beta)$ in the training set, with the simulation training data histogrammed in black. The bottom row plots the flow output for IMF parameters outside the training set. As expected, the mass spectra generally have larger uncertainties at interpolated values.}
\label{fig:flow_example}
\end{figure*}

In Figure~\ref{fig:flow_example}, we compare the training simulations to example predictions for the Pop~III primary mass distributions and their corresponding uncertainties from the flow ensemble. The emulator accurately reproduces the simulated populations interpolated across the different IMFs and the ensemble predicts uncertainty bands comparable to the histogram noise at each IMF location. We also show predictions for new IMFs not available in the training set where, crucially, the ensemble reliably produces larger prediction uncertainties.

\subsubsection{Pop~III star-formation history: nonparametric}\label{sec:p3sfrd}

By adopting models for the time delay between star formation and BH coalescence as well as the merger efficiency, we can infer the Pop~III SFRD directly instead of the MRD \cite{Fishbach:2024hfw}.
The source-frame MRD at redshift $z(t_s)$ corresponding to a source frame time $t_s$ is
\begin{align}
\mathcal{R} \big( z(t_s) \big)
=
\int_0^{\tau_\mathrm{max}} \dd{\tau}
\psi(t_s-\tau) \, p(\tau)\, \eta
\, ,
\end{align}
where $\tau$ is the delay time between formation and BH merger, $\psi(t_s-\tau)$ is the SFRD at the formation time $t_s-\tau$, $p(\tau)$ is the time-delay distribution, and $\eta$ is the merger efficiency; i.e., the number of BBHs that merge per solar mass of star formation.

We model the SFRD as a piecewise linear function with a Gaussian process~\cite[GP,][]{mackay2003information, williams2006gaussian} prior on the heights of the nodes. A GP is a stochastic process for which any finite collection of function values follow a multivariate normal distribution, which allows us to impose a prior assumption that the SFRD is a smooth function over redshift without using a stringent functional form. For computational efficiency, we define the process at a small number of redshifts; we found that five evenly spaced values between $z=10$ and $z=22$ were sufficient to describe a range of SFRDs including the TOP and LOG SFRDs we used as our underlying astrophysical populations. Specifically, the GP defines a prior on $\ln\psi_i$ for each of the five redshifts $z_i$ ($i=1,...,5$), whose values we interpolate linearly. More details are provided in Appendix~\ref{app:gp}.

We treat the time-delay distribution $p(\tau)$ as known and model it as $p(\tau) \propto \tau^\gamma$
with a minimum time delay $\tau_\mathrm{min} = 3$ Myr and slope $\gamma = -1$, which match the predicted scaling~\cite{Dominik2012} and the distributions returned by the Pop~III simulations~\cite{Costa:2023xsz} well. Although we tested fitting for the time-delay parameters during inference, we found them to be largely degenerate with the flexible SFRD model.

While the merger efficiency $\eta$ nominally depends on metallicity $Z$, the simulations we use assumed all Pop~III star formation happens at the same metallicity $Z=10^{-11}$. Following \citet{Santoliquido2023}, we also assume a unity binary fraction when converting from star-forming mass to BH mergers. The merger efficiency in this case therefore depends solely on the initial conditions of the binary stars. As we only consider variations of the IMF parameters $\alpha$ and $\beta$, as described in Sec.~\ref{sec:popIIIstars}, we thus interpolate the efficiencies $\eta$ derived from the four input simulations across the IMF parameter space and use the predicted $\eta(\alpha,\beta)$ from the sampled IMF parameters during inference. 

\section{Results}
\label{sec:results}

\subsection{Distinguishing stellar populations}\label{sec:distinguishing}

\begin{figure*}[!htb]
\centering
\includegraphics[width=0.98\textwidth]{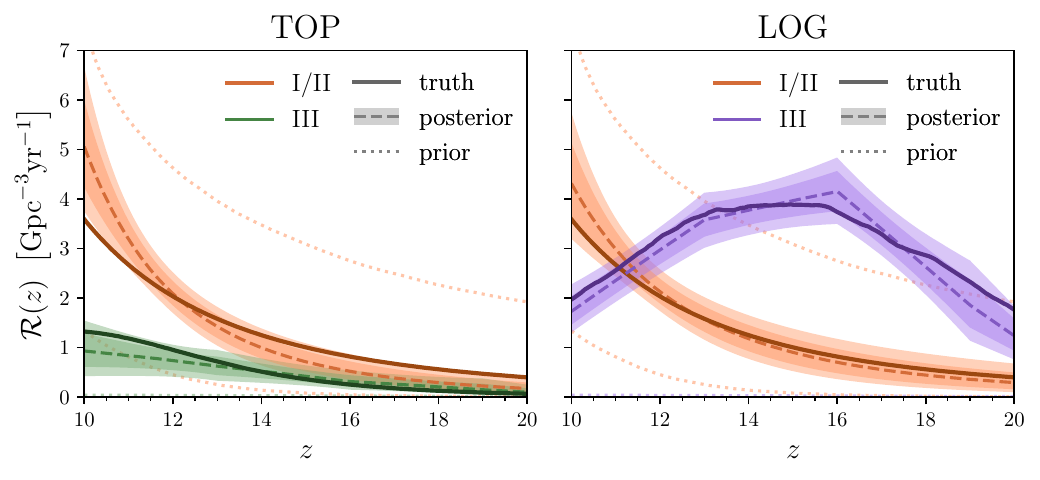}
\caption{Inferred source-frame merger rate densities. The left (right) panel shows the results for the TOP (LOG) Pop~III model. The Pop~I/II binaries are in \locclr in both panels, while TOP and LOG are in \TOPclr and \LOGclr, respectively. The dashed lines show the median inferred values. The colored bands span the 68\% and 90\% symmetric credible intervals. The true MRDs are plotted in dark solid lines and the dotted lines denote the central 90\% prior range for each model; the 95\% prior limit for the Pop~III merger rate density extends beyond the plot range.}
\label{fig:MRDs}
\end{figure*}

As we expect the formation histories for Pop~I/II and Pop~III binaries to differ, the BBH merger rate density is a key feature for discriminating between these populations. We show the inferred MRDs for each model in Figure~\ref{fig:MRDs}. At all redshifts, we exclude $\mathcal R_\mathrm{III}=0$ at nearly 100\% credibility; further, the inference is constrained away from the prior lower bound. Specifically, we measure the TOP MRD at $z=10$ to be $\mathcal R_\mathrm{III}(z=10)=\TOPMRDten \, \mathrm{Gpc}^{-3}\mathrm{yr}^{-1}$, distinguished from the Pop~I/II MRD, measured to be $\mathcal R_\mathrm{I/II}(z=10)=\TOPlocMRDten \, \mathrm{Gpc}^{-3}\mathrm{yr}^{-1}$. Analogously, we measure the LOG MRD $\mathcal R_\mathrm{III}(z=10)=\LOGMRDten \, \mathrm{Gpc}^{-3}\mathrm{yr}^{-1}$ alongside the Pop~I/II $\mathcal R_\mathrm{I/II}(z=10)=\LOGlocMRDten \, \mathrm{Gpc}^{-3}\mathrm{yr}^{-1}$.

\begin{figure*}[!htb]
\centering
\includegraphics[width=0.9\columnwidth]{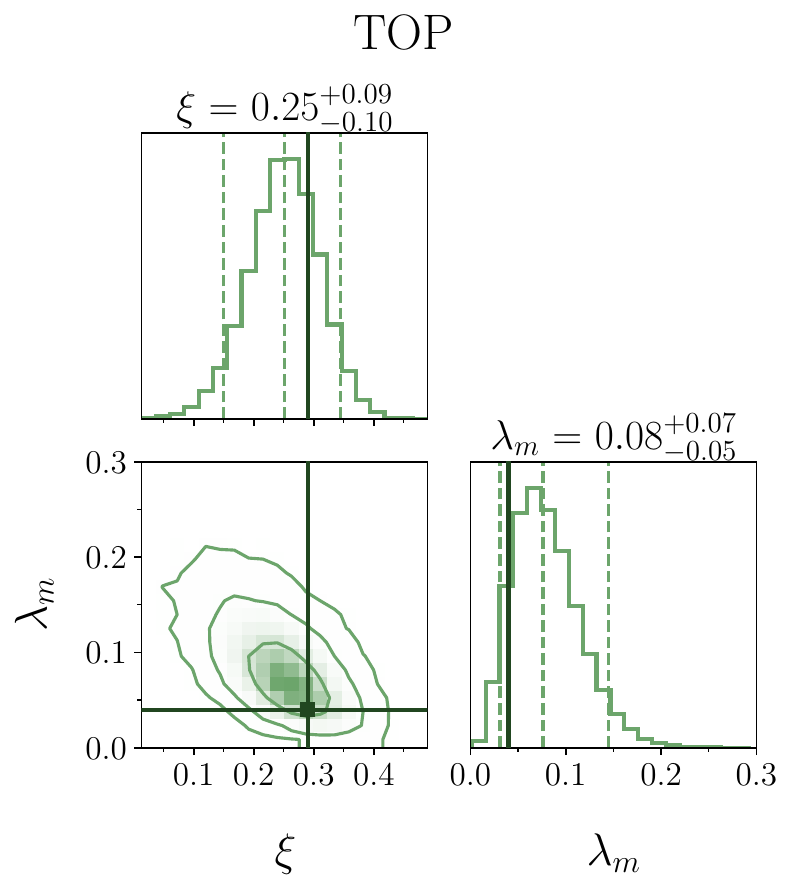}
\hspace{1cm}
\includegraphics[width=0.9\columnwidth]{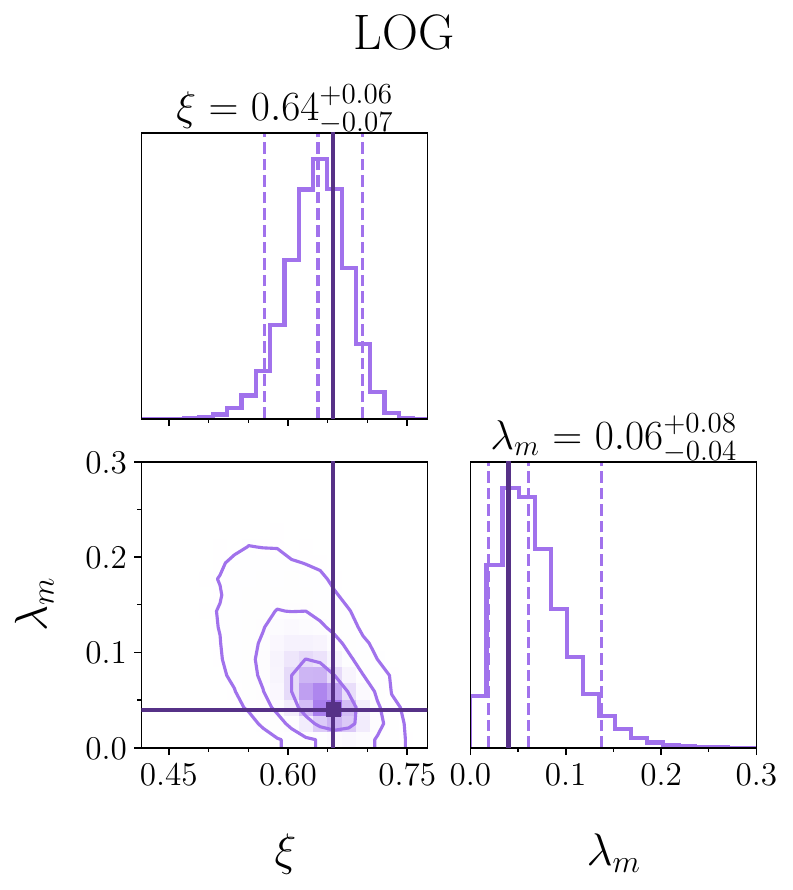}
\caption{Inferred Pop~III branching ratio $\xi$ and fraction of events in the Pop~I/II \textsc{Power Law + Peak} peak $\lambda_m$ for TOP and LOG. The contours encircle the 50\textsuperscript{th}, 95\textsuperscript{th}, and 99\textsuperscript{th} percentiles. The column labels give the median and 90\% credible intervals, also shown by dashed lines on the one-dimensional histograms.}
\label{fig:branch_peak_corr}
\end{figure*}

The LOG MRD is well-measured and distinguished from the I/II mergers. However, the posterior on the TOP MRD is systematically low at $z\sim10$. In turn, the Pop~I/II MRD is slightly overestimated, with the truth $\mathcal R_{10}=3.59$ falling at the $\sim3\%$ credible level, suggesting some Pop~III events are erroneously attributed to the local population. To test this association, we look for sampling correlations between the III--I/II branching ratio and other population parameters. While the Pop~III branching ratio $\xi$ is not an explicit hyperparameter, it can be obtained by integrating each detector-frame merger rate $R(z,\theta|\Lambda)$ (Eq.~\ref{eq:MRD}) over redshift and binary parameters $\theta$ to obtain the rate $dN/dt$ for each subpopulation. Indeed, as shown in Figure~\ref{fig:branch_peak_corr}, in both universes we find a negative correlation between the inferred fraction of Pop~III binaries and the \textsc{Power Law + Peak} peak fraction, $\lambda_m$, with nearly-identical Pearson correlation coefficients of $-0.54$ and $-0.53$ for TOP and LOG, respectively. Because both Pop~III populations are heavily weighted toward masses $M\gtrsim30M_\odot$, and in fact have local peaks near $35 M_\odot$ (see the two upper left panels of Fig.~\ref{fig:flow_example}), this implies some of the high-mass Pop~III events can be incorrectly assigned to the Gaussian peak of the local population. The smaller catalog size (215 events) and qualitatively similar shape of the MRDs likely contribute to the weaker discrimination between the TOP and Pop~I/II populations vis-à-vis the LOG case, in which the correlation persists but the MRDs and branching ratio are better-measured.

In the results presented in Fig.~\ref{fig:MRDs}, we use the parametric Pop~I/II MRD model described in Sec.~\ref{sec:locmodels}. However, the high-redshift evolution of the local population is highly uncertain and our choice of a power-law MRD is a strong modeling assumption. To test if we can distinguish the populations with weaker assumptions about their redshift evolution, we also run inference using a nonparametric model for the Pop~I/II MRD. Using a similar piecewise linear model with a GP prior as for the Pop~III SFRD (Sec.~\ref{sec:p3sfrd} and App.~\ref{app:gp}), we find we can still differentiate the populations' MRDs albeit with weaker constraints. These results imply our analysis is robust to the precise modeling choice for the redshift evolution of the local binaries.

\subsection{Pop~III constraints}

\begin{figure*}[!htb]
\centering
\includegraphics[width=0.98\textwidth]{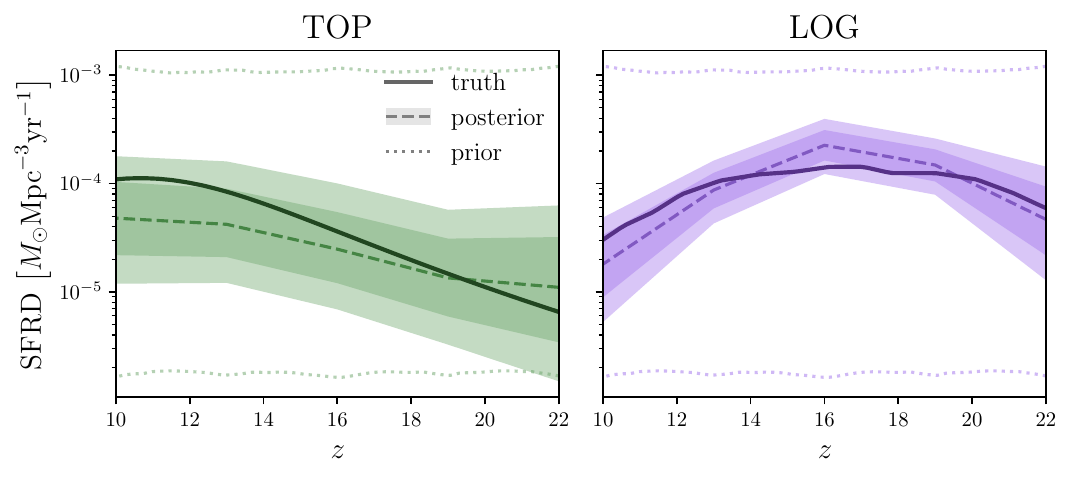}
\caption{As in Fig.~\ref{fig:MRDs}, but for the star formation rate densities for the two Pop~III universes.}
\label{fig:SFRDs}
\end{figure*}

As described in Sec.~\ref{sec:p3sfrd}, we did not directly model the Pop~III MRD, but instead the SFRD, from which we obtained the MRD by convolving with a log-uniform time delay distribution and  merger efficiency given by the sampled IMF parameters.
Fig.~\ref{fig:SFRDs} shows our inference results on the SFRD for each Pop~III model. The lines and colors are akin to those in Fig.~\ref{fig:MRDs}. For both Pop~III models, we correctly infer the SFRD, although the constraints for LOG are much tighter owing to the $\sim5$-fold increase in Pop~III catalog size. The uncertainty on the SFRD does not significantly evolve with redshift. For the TOP (LOG) model, we measure $\ln\ \psi/(M_\odot\rm{Mpc}^{-3}\rm{yr}^{-1})$ to be \TOPSFRDten (\LOGSFRDten) at $z=10$, and \TOPSFRDsixteen (\LOGSFRDsixteen) at $z=16$.

As we use a fixed time-delay model, $p(\tau)\propto\tau^{-1}$, our SFRD inference results are contingent on the chosen time-delay model being a good approximation to the truth. Here, we have access to samples from the true time-delay distributions predicted by the BPS simulations, so we are able to confirm that a log-uniform is a reasonable fit to the data. If we used an incorrect time-delay model---at least, one further from the truth---we would get biased inference on the SFRD. However, as mentioned in Sec.~\ref{sec:p3sfrd}, fitting for the time-delay parameters is impractical with a flexible SFRD model. While modeling the SFRD parametrically would alleviate the strong degeneracy with the time-delay distribution, it would in turn remove the benefits of the flexible model.

\begin{figure*}[!htb]
    \centering
    \includegraphics[width=0.9\columnwidth]{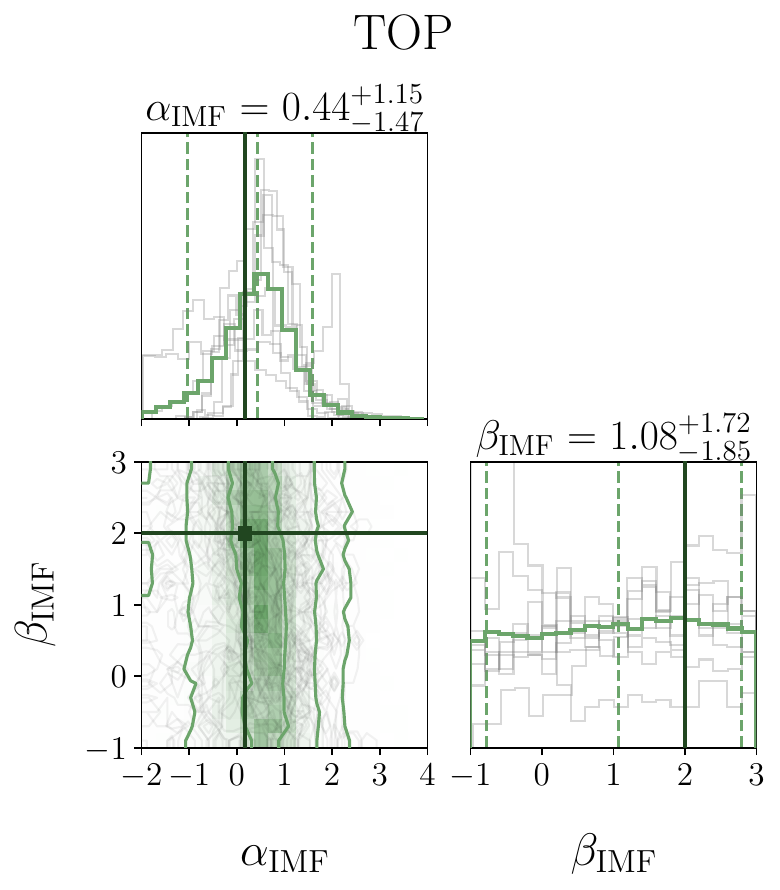}
    \hspace{1cm}
    \includegraphics[width=0.9\columnwidth]{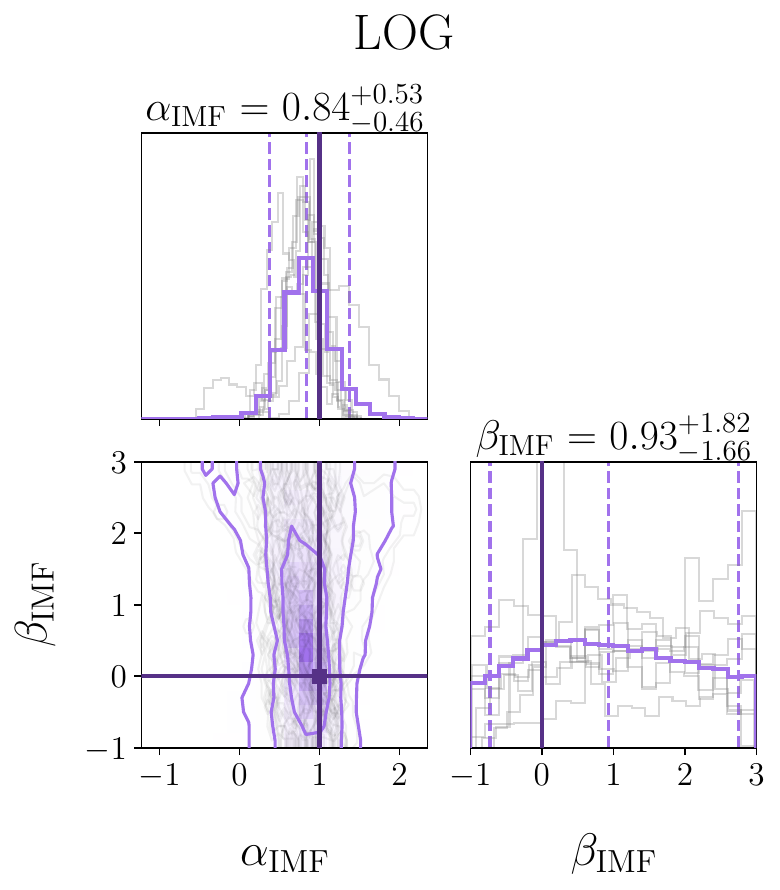}
    \caption{Posteriors on the IMF parameters for the TOP (left) and LOG (right) Pop~III models. Overplotted in gray are the inference results from eight flows in the ensemble to demonstrate the effect of training uncertainty prior to marginalization. The contours encircle the 50\textsuperscript{th}, 95\textsuperscript{th}, and 99\textsuperscript{th} percentiles. The column labels provide the medians and 90\% credible intervals, also denoted by dashed lines on the histograms.}
    \label{fig:IMFs}
\end{figure*}

The inferred IMF parameters for each model are in Figure~\ref{fig:IMFs}. Overplotted in light gray lines are the separate inference results from eight of the flows in the ensemble, all 50 of which we then combine with equal weighting to obtain the marginalized posterior, as covered in Sec.~\ref{sec:p3mm}. The differences between individual flow posteriors and the marginalized posterior demonstrate that an ensemble is necessary to correctly characterize the uncertainty in the IMF parameter posteriors. To test the effect of the ensemble size, we compare the ensemble-marginalized posterior (Eq.~\ref{eq: ensemble posterior}) using different numbers of flows $W$ between 1 and 50. We find that $W\sim10$ flows are sufficient to describe the posterior; the difference between the $W=n$ and $W=50$ cases diminishes rapidly for $n\gtrsim10$.

While the power-law index of the IMF, $\alpha_\mathrm{IMF}$, is constrained from the prior for both models, the exponential index $\beta_\mathrm{IMF}$ is poorly measured. For TOP (LOG), we find the median and 90\% credible intervals on the IMF parameters to be $\alpha_{\rm{IMF}}=\TOPIMFalpha \ (\LOGIMFalpha)$ and $\beta_{\rm{IMF}}=\TOPIMFbeta \ (\LOGIMFbeta)$. The weak constraints on $\beta_\mathrm{IMF}$ arise from the fact that the BPS simulations used to train the model are not as sensitive to changes in $\beta_\mathrm{IMF}$; the power-law index $\alpha_\mathrm{IMF}$ has a stronger impact on the distributions. In fact, as seen in Figure~\ref{fig:flow_example}, the primary mass  distribution $p(m_1)$ is nearly identical for the KRO and LAR models, which have nearly the same $\alpha_\mathrm{IMF}$ but differ in $\beta_\mathrm{IMF}$. The effect of the different $\beta_\mathrm{IMF}$ parameter primarily manifests in a change in merger efficiency. Consequently, the normalizing flows do not predict strong evolution with $\beta_\mathrm{IMF}$ at fixed $\alpha_\mathrm{IMF}$ and $\beta_\mathrm{IMF}$ is weakly measured during inference.

Because the SFRD-to-MRD convolution depends on the merger efficiency computed from the IMF parameters, it is plausible that the IMF inference is driven by the need to match the scale of the MRD, rather than the shape of the mass distribution. To test whether the IMF inference depends on its importance in the MRD computation, we repeat inference for five of the normalizing flows in the ensemble with the merger efficiency fixed to its true value. We find the variances in the IMF parameter posteriors are nearly identical to the variances when $\eta$ is a function of the sampled IMF parameters, although the constraints on the SFRD are much tighter since the efficiency, which is a dominant factor in setting the MRD scale, is fixed. This implies the IMF inference is not dependent on its use in the MRD convolution; conversely, uncertainty in the SFRD inference is in part driven by the uncertainty in the mass distribution.

\begin{figure*}[!htb]
\centering
\includegraphics[width=0.98\textwidth]{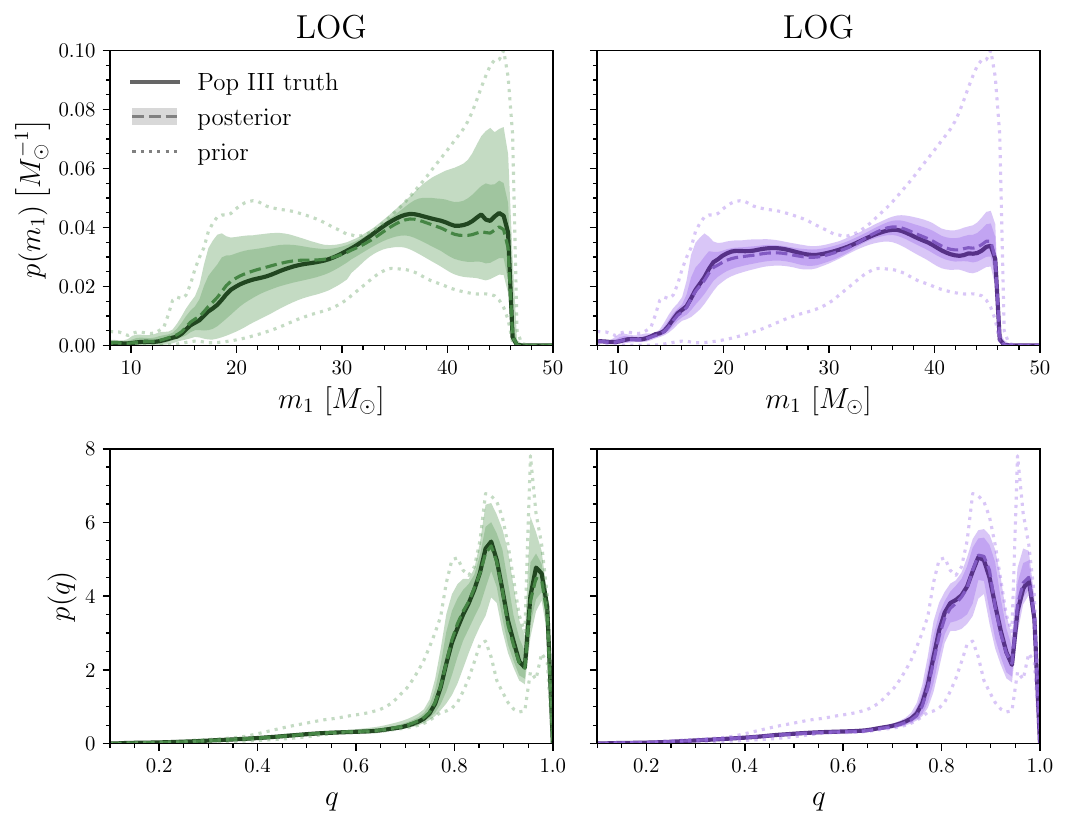}
\caption{Similar to Fig.~\ref{fig:MRDs}, but for the predictive distributions for Pop~III masses. The top row shows the primary mass distribution marginalized over mass ratio and redshift; the bottom row shows the mass ratio distribution marginalized over primary mass and redshift.}
\label{fig:III_m1}
\end{figure*}

\begin{figure*}[!htb]
    \centering
    \includegraphics[width=0.98\textwidth]{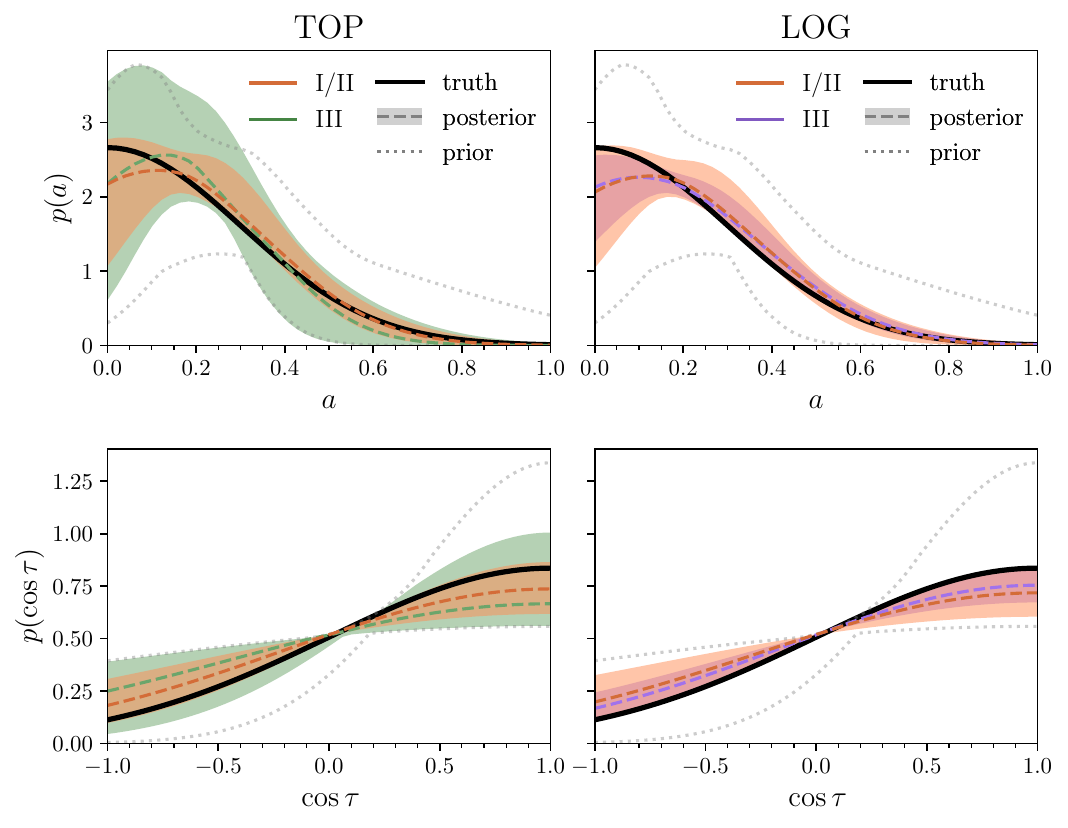}
    \caption{Inferred spin magnitude (top panel) and cosine tilt (lower panel) distributions. The figure elements are as in Fig.~\ref{fig:MRDs}, except the shared truth is in black and for clarity we do not plot the 68\% credible regions.}
    \label{fig:spinstilts}
\end{figure*}

We also show the posterior predictive distributions for the Pop~III primary mass in Figure~\ref{fig:III_m1}, marginalized over the inferred redshift distribution and mass ratio.
The inferred spin and tilt distributions for each universe are shown in Figure~\ref{fig:spinstilts}. The top panel displays the inference for the dimensionless spin magnitude $a$ and the lower panel for the cosine tilt angle $\cos\tau$. In both cases, we constrain the spin and tilt distributions away from the prior; the LOG spins are better measured than the TOP spins, again owing to the larger catalog size.

\subsection{Pop~I/II constraints}

\begin{figure*}[!htb]
\centering
\includegraphics[width=0.98\textwidth]{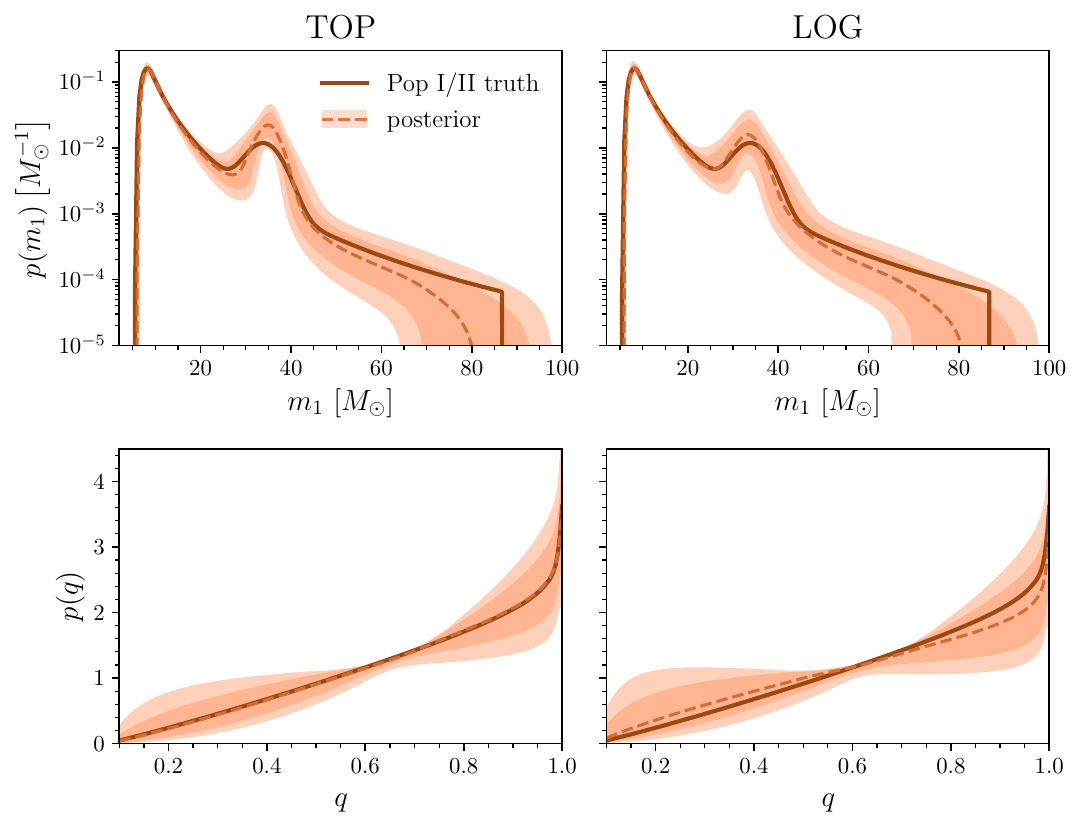}
\caption{Posterior predictive distributions for the local, Pop~I/II mass distribution. The figure elements are as in Fig.~\ref{fig:MRDs}. The top panel shows the primary mass while the lower panel shows the mass ratio distribution.
}
\label{fig:local_m1}
\end{figure*}

Simultaneously with our inference on the properties of the Pop~III stars, we recover those of the Pop~I/II binaries. The inferred mass distribution for the local subpopulation is shown in Figure~\ref{fig:local_m1}. The top panel shows the primary mass while the lower panel plots the mass ratio inference. Fig.~\ref{fig:spinstilts} shows the inference on the Pop~I/II spin and tilt distributions, for which we use the same model as for the Pop~III binaries but use independent parameters. The mass and spin distributions are marginalized over the flow ensemble. Unlike the Pop~III IMF inference, the Pop~I/II parametric inference is minimally affected by the choice of normalizing flow; the variation between individual flow posteriors is negligible compared to the variance within each posterior.

\section{Conclusion}\label{sec:conclusion}

BBHs from Pop~III stars are an exciting possible source for next-generation gravitational-wave observatories, which may be able to detect and accurately measure the parameters of stellar-mass BH binaries out to redshifts $z\gtrsim20$. Since GW sources are not labeled with their origins, extracting the astrophysical properties of Pop~III stars from GW data requires models for their stellar properties as well as for the existence and features of subpopulations in the data.

We used a mixture of nonparametric, parametric, and simulation-based modeling approaches to infer the properties of Pop~III stellar populations in a universe that also contained Pop~I/II binaries. By doing so, we leveraged the benefits of each approach and encoded different strengths of assumptions for each component of the population. In particular, we modeled the Pop~III star formation rate density nonparametrically and used a surrogate model trained on binary population synthesis simulations to infer the Pop~III initial mass function. We developed a method to account for uncertainty in the emulator, demonstrated it correctly calibrated the model and posterior, and showed we can perform inference even with a limited number of training simulations. With our novel mixed-modeling method, we demonstrated we can distinguish the Pop~III from Pop~I/II binaries even at redshifts where the subpopulations have comparable rates, despite the uncertainty of the surrogate and the flexibility of the merger rate density model. We found we can constrain the Pop~III star formation history and spectral index of the initial mass function with one year of data, although the exponential index of the initial mass function was not well-measured. In the next-generation era, such constraints on the Pop~III progenitor population will contribute to a better understanding of star formation in the early universe, helping untangle remaining questions about the processes of reionization and the chemical enrichment of the first galaxies.

However, our results bear caveats due to our remaining modeling assumptions. We assumed we knew the population comprised two parts and that the parametric and simulation-based recovery models had the correct parameterizations, benefits not afforded to inference on real data. We supposed the local population was well-characterized by a known parametric model and that Pop~III binaries closely matched the expectation from the astrophysical simulations. While this is similar to what is done in most of the literature, we should stress that missing components of the population can lead to biased inference~\cite{Cheng:2023ddt} and it is unlikely that Pop~III BBH precisely follow the distributions predicted by current population synthesis methods. Interpreting the astrophysical origins of sources in next-generation detectors should still be done with caution.

This said, over the next decade, theoretical models for Pop~III stars will continue to be informed by indirect observational constraints, binary population-synthesis modeling will include increasingly detailed stellar-evolution tracks, and the features of the local binary black-hole population will be measured with higher precision by the current generation of detectors~\cite[e.g.,][]{Hartwig:2024wnk, Koutsouridou:2023ibj, Andrews:2024saw, T2200287}. Model selection and\//or residual tests can be used to at least exclude population-synthesis models whose predictions are at odds with the observed data. Our surrogate training method can be extended to support further simulations of Pop~III stars and our analysis framework can accommodate other subpopulations or different models for the local population, though at higher computational expense and potentially degraded constraints on some parameters. The risk of bias introduced by models that are too rigid can be reduced by allowing for subpopulations described using nonparametric approaches~\cite[e.g.,][]{Edelman:2022ydv, Ray:2023upk, Tiwari:2020vym,Heinzel:2024jlc}. Depending on the degree of elasticity with which one endows the nonparametric part of the model, correlations within and between subpopulations might inflate the uncertainties.

Further, it is worth remembering that our results are based on just one year of simulated data: larger catalogs will yield smaller statistical uncertainties. The approach we presented introduces promising ways of extracting astrophysical information from gravitational-wave catalogs, highlights the benefits of integrating modeling paradigms, and reinforces the science case for next-generation gravitational-wave observatories as probes of the early universe.

\acknowledgments

We thank Sofía Álvarez-López, Tom Callister, Storm Colloms, Maya Fishbach, Jack Heinzel, Michela Mapelli, Filippo Santoliquido, and Noah Wolfe for useful discussions.
C.P. is partially supported by the NSF grant PHY-2045740.
S.V. is partially supported by the NSF through grants PHY-2045740, PHY-2309064 and PHY-2207387.
M.M. is supported by the LIGO Laboratory through the National Science Foundation awards PHY-1764464 and PHY-2309200.
The authors are grateful for computational resources provided by subMIT at MIT Physics and the LIGO Laboratory supported by National Science Foundation Grants PHY-0757058 and PHY-0823459.

\appendix

\section{Normalizing flows}
\label{app:flows}

Normalizing flows are a type of generative model that can emulate high-dimensional conditional probability distributions, providing both efficient sampling and exact density evaluation \cite{JMLR:v22:19-1028, 9089305}.
They consist of a sequence of invertible, differentiable transformations from a simple latent space with a known base distribution (such as an multivariate standard normal distribution) to a complex target distribution and are trained by optimizing the transformations to best match the target. A vector $\vec x$ in the target space is expressed as a transformation $f$ of a vector $\vec z$ in the latent space sampled from the base distribution $p_z$:
\begin{equation}
\vec x
=
f(\vec z)
\, , \quad
\vec z\sim p_z
\, .
\end{equation}
For invertible, differentiable transformations $f$, the density of $\vec x$ is related to that of $\vec z$ by
\begin{equation}
    p_x(\vec x) = p_z \big( f^{-1}(\vec x) \big) \big| \det J_f(\vec z) \big|^{-1}
,
\end{equation}
where
\begin{equation}
    \det J_f(\vec z) = \left|\frac{\partial f(\vec z)}{\partial \vec z}\right|
\end{equation}
is the Jacobian determinant of $f$. While finding a single bijection that can produce an arbitrarily complex distribution can be difficult, one can exploit the property that the composition of bijective functions is also bijective. For a series of transformations $f_1,\dots,f_n$ starting from a sample $\vec z_0$ from the base distribution $p_z$, the composition $f = f_1\circ\cdots\circ f_n$ has Jacobian determinant $\det J_f(\vec z_0) = \prod_{i=1}^n\det J_{f_i}(\vec z_{i-1})$, where $\vec z_i=f_i(\vec z_{i-1})$. The relation between the target and latent densities is then
\begin{equation}
p_x(\vec x)
=
p_z(\vec z_0) \prod_{i=1}^n \left| \det \frac { \partial f_i(\vec z_{i-1}) } { \partial \vec z_{i-1} } \right|^{-1}
.
\end{equation}
Linking a chain of transformations therefore allows one to construct increasingly complex densities $p_x(\vec x)$.

That normalizing flows can produce flexible densities makes them useful for emulation. Suppose we have a parameter $\theta$ (e.g., BBH mass) that conditionally depends on another parameter $\Lambda$ (e.g., IMF slope). Consider a set of samples $\{\theta_1,\dots,\theta_n\}$ and associated  $\{\Lambda_1,\dots,\Lambda_n\}$ drawn from the unknown underlying distribution $p(\theta|\Lambda)$; i.e., $\theta_i\sim p(\theta|\Lambda_i)$. We aim to approximate the unknown $p(\theta|\Lambda)$ with $q(\theta|\Lambda,\omega)$, where $q$ is the distribution predicted by a normalizing flow and $\omega$ are its parameters, e.g., the weights of a neural network. To do so, we optimize the set of flow transformations $f$ by maximizing the likelihood $p(\mathcal{T}|\omega) \propto \prod_{i=1}^n q(\theta_i|\Lambda_i)$ that the training data $\mathcal{T} = \{\theta_i,\Lambda_i\}$ are IID draws from $q$. In practice, we equivalently minimize the cross-entropy 
\begin{equation}
L =
- \frac{1}{n} \sum_{i=1}^n 
\log q(\theta_i|\Lambda_i)
\, .
\end{equation}

We use the \textsc{FlowJAX}~\cite{ward2023flowjax} implementation of coupling layers~\cite{Dinh2016} with rational quadratic splines~\cite{Durkan2019} to transform from a standard-normal base distribution to our target BBH populations. As with many numerical methods, normalizing flows work optimally with parameters on roughly unit scales. To match the supports of the standard normal and BBH parameters, we first use the standard logistic function followed by affine transformations to unbound and rescale each physical parameter---$m_1, q,z,\alpha,\beta$--to zero mean and unit variance. In each flow layer, the parameter vector $\theta$ is split into two halves: the first is transformed by the spline; the second undergoes the identity transformation and is input along with the parameters $\Lambda$ to a standard feedforward neural network that conditions the splines by outputting the knot locations, heights, and derivatives.
Between each layer, the parameter vector is randomly permuted.
For each flow, we use a sequence of three coupling layers. In each layer we use a conditioner network with a single hidden layer of 10 neurons and ReLU activation functions~\cite{2018arXiv180308375A}. For the spline transformations we use three knots.
This architecture was determined empirically as the best tradeoff between model complexity and computational efficiency without overfitting the training data. We train with Adam optimization~\cite{kingma2017} and a cosine-decay learning rate from $10^{-2}$ to $10^{-5}$.

\section{Population likelihood}
\label{app:likelihood}

Suppose we observe $N$ events with data $\mathcal D = \{d_1,\dots,d_N\}$ over a time $T$. We suppose the population is described by a set of hyperparameters $\Lambda$, which determine the rate density of mergers as a function of their event-level parameters $\theta$. We describe their detector-frame merger rate 
\begin{equation}
R(\theta|\Lambda) = \frac{\dd{N}}{\dd{t_\mathrm{d}} \dd{\theta} \dd{z}} \bigg|_\Lambda,
\end{equation}
where $N$ is the total number of mergers and $t_{\mathrm d}$ is detector-frame time. In practice, we model the differential comoving merger rate $\mathcal{R}(\theta|\Lambda)$, the number of mergers per unit comoving volume $V_\mathrm{c}$ and source-frame time $t_\mathrm{s}= (1+z)^{-1}t_\mathrm{d}$ under the population hyperparameters $\Lambda$, and convert to $R(\theta|\Lambda)$:
\begin{equation}
R(\theta|\Lambda)
=
\frac{\dd{t_\mathrm{s}}}{\dd{t_\mathrm{d}}}
\frac{\dd{V_\mathrm{c}}}{\dd{z}}
\frac{\dd{N}}{\dd{t_\mathrm{s}} \dd{\theta} \dd{V_c}}
=
\frac{1}{1+z} \frac{\dd{V_\mathrm{c}}}{\dd{z}} \mathcal{R}(\theta|\Lambda).
\end{equation}

Assuming GW events are the result of an inhomogeneous Poisson process, we can derive a total population likelihood \cite{Mandel:2018mve, Vitale2020}
\begin{equation}\label{eq:lkl}
    \mathcal{L}(\mathcal{D}|\Lambda) \propto e^{-N_\mathrm{exp}(\Lambda)} \prod_{n=1}^N \mathcal L_n(\Lambda),
\end{equation}
where the individual event likelihoods $\mathcal L_n$ are given by~\cite{Loredo2004}
\begin{align}
    \mathcal L_n(\Lambda) = R(d_n|\Lambda)=\int \dd{\theta_n} p(d_n|\theta_n) R(\theta_n | \Lambda)
\end{align}
and the expected number of detections under the population model $\Lambda$ is
\begin{align}
     N_\mathrm{exp}(\Lambda) = T \int \dd{\theta} p(\mathrm{det}|\theta)R(\theta | \Lambda).
\end{align}
Selection effects are encoded in the probability of detecting a signal with parameters $\theta$, $p(\mathrm{det}|\theta)$, which itself is an integral over possible data realizations $d$:
\begin{align}
     p(\mathrm{det}|\theta) = \int \dd{d} p(\mathrm{det}|d)p(d|\theta).
\end{align}

The individual event likelihoods and expected detections integrals are computationally intractable, so we typically approximate these integrals via importance sampling. By Bayes' theorem, for an individual-event posterior inferred under a default parameter estimation prior $\pi$, we can write $p(d|\theta)\propto p(\theta|d)/\pi(\theta)$ and rewrite the likelihood:
\begin{align}
    \mathcal L_n(\Lambda) \propto \int \dd{\theta_n} p(\theta_n|d_n)\frac{R(\theta_n | \Lambda)}{\pi(\theta_n)},
\end{align}
which we can then replace with a Monte Carlo estimator~\cite{Essick:2022ojx}
\begin{align}
\label{eq:Lhatn}
\Lhat_n(\Lambda) = \frac{1}{N_\mathrm{PE}} \sum_{k=1}^{N_\mathrm{PE}} \frac{R(\theta_{n,k}|\Lambda)}{\pi(\theta_{n,k})},
\end{align}
where $N_\mathrm{PE}$ is the number of samples in the initial parameter estimation posterior in which we assume each sample is a fair draw $\theta_{n,k}\sim p(\theta_n|d_n)$.

Similarly, we estimate selection effects $p(\mathrm{det}|\theta)$ and the expected detections integral using a reference set of $N_\mathrm{inj}$ injections drawn from a distribution $p_\mathrm{inj}(\theta)$, of which $N_\mathrm{det}$ are above threshold and detected~\cite{Tiwari:2017ndi}:
\begin{align}
    \hat N_\mathrm{exp}(\Lambda) = \frac{T}{N_\mathrm{inj}} \sum_{i=1}^{N_\mathrm{det}}\frac{R(\theta_i | \Lambda)}{p_\mathrm{inj}(\theta_i)}.
\end{align}

The total log-likelihood estimator is then
\begin{align}
\label{eq:logLhat}
\ln \Lhat(\mathcal{D}|\Lambda) \propto
- N_\mathrm{exp}(\Lambda)
+ \sum_{n=1}^N \ln \hat{\mathcal L}_n(\Lambda).
\end{align}
While likelihood estimators make the computation tractable, they introduce inherent statistical uncertainty due to the finite number of samples used in the Monte Carlo integrals. The variances in the estimators $\Lhat_n(\Lambda)$ and $\hat N_\mathrm{exp}(\Lambda)$ are, respectively,
\begin{align}
    \hat{\sigma}_n^2 &= \frac{1}{N_\mathrm{PE}^2} \sum_{k=1}^{N_\mathrm{PE}} \left(\frac{R(\theta_{n,k}|\Lambda)}{\pi(\theta_{n,k})}\right)^2 - \frac{\Lhat_n^2}{N_\mathrm{PE}},\\
    \hat{\sigma}_\mathrm{exp}^2 &= \frac{T^2}{N_\mathrm{inj}^2} \sum_{i=1}^{N_\mathrm{det}}\left(\frac{R(\theta_i|\Lambda)}{p_\mathrm{inj}(\theta_i)}\right)^2 - \frac{\hat N_\mathrm{exp}^2}{N_\mathrm{inj}}.
\end{align}
These combine to produce a total variance in the log-likelihood estimator of
\begin{align}\label{eq:lklvar}
    \hat{\sigma}_{\ln\Lhat}^2 = \sum_{n=1}^N \frac{\hat{\sigma}_n^2}{\Lhat_n^2} + \hat{\sigma}_\mathrm{exp}^2.
\end{align}
If the variance is large, the estimate of the log-likelihood in Eq.~(\ref{eq:logLhat}) is not trustworthy, though the relevant threshold is a subject of debate. Ongoing research concerns the appropriate procedure to handle variance in the likelihood estimator \cite{Farr:2019rap, Essick:2022ojx, Talbot:2023pex}. Here, we follow \citet{Heinzel:2024jlc} and effectively impose a variance cut of $\hat{\sigma}_{\ln\Lhat}^2\lesssim1$ via a steep taper on the likelihood function, $\ln\Lhat \to \ln\Lhat - \mathcal T$, with the penalty
\begin{align}
\mathcal T =
\begin{cases}
100\left(\hat{\sigma}_{\ln\Lhat}^2-1\right)^2
&\mathrm{if} \ \hat{\sigma}_{\ln\Lhat}^2 \geq 1, \\
0
&\mathrm{if} \ \hat{\sigma}_{\ln\Lhat}^2 < 1.
\end{cases}
\end{align}

We use our catalogs of detected events and their reference posteriors (Section~\ref{sec:catalogs}) to sample the population posteriors. Although we generate events only within $z\in[10,20]$, we set a uniform prior on luminosity distance that allows for samples above and below the true event ranges. This is because noise can ``promote" (demote) the SNR of an event so its redshift posterior is systematically close (far). Forcing the redshift posteriors of these events to lie within $z\in[10,20]$ biases the inference on mass and spin, which affect the shape and scale of the signal similarly to redshift. However, because our population model places zero likelihood outside $z\in[10,20]$, events with a high fraction of posterior samples outside this range end up having large PE uncertainty. To avoid these events affecting the likelihood variance, we discard events for which more than 75\% of redshift samples are outside $[10,20]$. Although cutting catalog events in this manner introduces a small bias in population inference~\cite{Essick:2023upv}, we find it is a subdominant effect.

Specifically, we sample the population posteriors using the No-U-Turn-Sampler (\textsc{NUTS})~\cite{Hoffman2011} as implemented in \textsc{numpyro}~\cite{Phan2019}. We leverage the \textsc{JAX} framework for automatic differentiation, just-in-time compilation, and GPU acceleration~\cite{jax2018github}. We run inference separately for each normalizing flow in our ensemble (Appendix~\ref{app:flows}) and combine with equal weighting (Eq.~\ref{eq: ensemble posterior}) to marginalize over uncertainty in the neural network emulator.

\section{Nonparametric SFRD}
\label{app:gp}

We set a fixed mean of $\ln\psi/(M_\odot\mathrm{Mpc}^{-3}\mathrm{yr}^{-1})=-10$ for the GP and generated the covariance matrix for $\ln\psi_i$ with a squared-exponential kernel,
\begin{align}
k(z,z') =
\sigma^2 \exp \left( -\frac{(z-z')^2}{2\ell^2} \right)
\,,
\end{align}
evaluated on the input $z_i$ with an overall variance $\sigma^2=4$ and redshift length scale $\ell=6$. Rather than fit for these parameters, we heuristically fix them to reasonable values to alleviate some computational cost. Our choices encode prior assumptions about the smoothness of the SFRD. We find that changing or fitting for the GP parameters hardly affects the inference and the latter slows the sampling rate.

To sample from the GP, we first draw from independent standard normals for each redshift value $z_i\in[10,13,16,19,22]$. To transform into SFRD space, we multiply the standard normal draws by the covariance matrix rotated using the Cholesky decomposition then add the overall mean of $-10$.

\bibliographystyle{apsrev4-1}
\bibliography{refs}

\end{document}